\documentclass[12pt]{article}
\usepackage{amsmath}
\usepackage{amssymb}
\usepackage{epsfig}
\usepackage{amsthm}

\numberwithin{equation}{section}
\numberwithin{figure}{section}

\def\eq#1{(\ref{eq:#1})}
\def\lineup{\!\!\!\!\!\!\!\!&&}

\newcommand{\Tr}{\mathop{\rm Tr}\nolimits}
\def\d{\partial}

\def\ci{\mathrm{ci}}
\def\si{\mathrm{si}}

\def\fraction#1#2{ { \textstyle \frac{#1}{#2} }}
\def\half{\fraction{1}{2}}

\def\eps{\epsilon}
\def\V{\mathcal{V}\,}

\def\be{\begin{eqnarray}}
 \def\ee{\end{eqnarray}}
 \def\0{\nonumber}

\begin{document}

\begin{titlepage}

\begin{center}

\vskip 1.0cm {\large \bf{Comments on Lumps from RG Flows}}
\\
\vskip 2.0cm

{\large Theodore Erler\footnote{Email: tchovi@gmail.com}}
\vskip .5cm

{\large Carlo Maccaferri\footnote{Email: maccafer@gmail.com}}

\vskip 1.0cm

{\it {Institute of Physics of the ASCR, v.v.i.} \\
{Na Slovance 2, 182 21 Prague 8, Czech Republic}}

\vskip 1.0cm
{\bf Abstract}
\end{center}
\noindent In this note we investigate the proposal of Ellwood\cite{Ian} and
one of the authors {\it et al}\cite{BMT} to construct a string field
theory solution describing the endpoint of an RG flow from a
reference $\mathrm{BCFT}_0$ to a target $\mathrm{BCFT}^*$. We show that the
proposed class of solutions suffers from an anomaly in the equations of
motion. Nevertheless, the gauge invariant action exactly reproduces the
expected shift in energy.
\noindent
\medskip

\end{titlepage}

\newpage

\tableofcontents

\baselineskip=18pt

\section{Introduction}

One of the major outstanding problems in open string field theory has been
to find an analytic solution describing a lower dimensional brane
(a tachyon ``lump'') from the perspective of a higher dimensional brane.
Following Sen's conjectures\cite{Sen,Sen2}, such a solution was constructed
numerically in the Siegel gauge level expansion by Moeller, Sen, and
Zwiebach\cite{Moeller}. An exact description of the tachyon lump,
however, has remained elusive.

One of the few concrete proposals for constructing the lump
was suggested by Ellwood\cite{Ian} in his analysis of the gauge structure
of open string field theory around the tachyon vacuum. Later Bonora,
Tolla, and one of the authors (henceforth BMT)\cite{BMT}, interpreted
Ellwood's proposal as a prescription which, given a boundary
RG flow which interpolates from a reference boundary conformal field theory
$\mathrm{BCFT}_0$ to a target boundary conformal field theory
$\mathrm{BCFT}^*$\cite{Ludwig}, produces a formal solution in
$\mathrm{BCFT}_0$ describing a configuration of branes corresponding to
$\mathrm{BCFT}^*$. The construction is fairly general, and includes formal
solutions for a single lump as a special case. BMT gave a simple
example of such a solution, and proved that it has the correct coupling to
closed string states\cite{BMT}.

In this paper we show that the Ellwood and BMT solutions, as currently
understood\footnote{As we will see, the Ellwood and BMT solutions are singular,
and it is possible that a nontrivial regularization exists which solves the
equations of motion. We will refer to the Ellwood and BMT solutions as
``solutions,'' though, as currently understood, they do not satisfy the
equations of motion.}, do not satisfy the equations of motion.
Carefully evaluating the equations of motion produces an anomalous, nonzero
state, proportional to a projector of the star algebra. The anomaly can be
interpreted as saying that the equations of motion are solved
only with respect to
the states of the IR boundary conformal field theory. Surprisingly,
we find that, nevertheless,
the action evaluated on the BMT and Ellwood solutions exactly accounts for
the total shift in the energy between the UV and IR boundary conformal field
theories. This is independent of the particular relevant boundary interaction
and depends only on universal properties of the RG flows we consider.

This paper is organized as follows. In section \ref{sec:review} we review
the algebraic setup and necessary
assumptions about the relevant boundary interaction.
In section \ref{sec:Ell_BMT} we introduce the Ellwood
and BMT solutions, and explain in general terms why they are singular
and why they are not expected to satisfy the equations of motion.
In section \ref{sec:BMT} we study the BMT solution in a regularization which
expresses it as a sum of a tachyon vacuum solution plus a ``phantom term''
which builds the IR boundary conformal field theory on top of the
tachyon vacuum. We show that the BMT solution does not satisfy the equations
of motion, either when contracted with the solution or with Fock space
states, but reproduces the correct difference in energy between the
perturbative vacuum and the boundary conformal field theory in the infrared.
In section \ref{sec:Ellwood} we extend these results to the (more general)
Ellwood solutions, where a few new complications arise.
In section \ref{sec:cohomology} we investigate whether
there is a sense in which the BMT solution supports the correct cohomology
of open string states. We argue that the equations of motion are
satisfied with respect to a class of projector-like states which
can be put into one-to-one correspondence
with the states of the IR boundary conformal field theory. Within
this subclass of states, the kinetic operator around the BMT solution
is nilpotent and its cohomology precisely corresponds to
on-shell states in the infrared. We end with some concluding remarks.\\

\noindent {\bf Note Added}:  While this paper was in preparation we were
notified of the work of \cite{BT} which contains some overlap with our
results. The papers should appear concurrently.

\section{Setup}
\label{sec:review}

In this section we review the basic ingredients needed to understand the
Ellwood and BMT solutions. We use the same setup as \cite{BMT}, but
add a few clarifications.

\subsection{RG Flows}
\label{subsec:RG}

The construction begins with a relevant matter boundary
operator, $\phi(s)$, which triggers an RG flow from a reference boundary
conformal field theory, $\mathrm{BCFT}_0$, to a target boundary conformal
field theory, $\mathrm{BCFT}^*$. For the string field theory manipulations we
need to perform, we need to assume that $\phi(s)$ satisfies three properties:
\begin{description}
\item{\bf 1)} {\it The $T$-$\phi$ OPE is no more singular than a double
pole.} This means
\begin{equation}T(s)\phi(0)
\sim \frac{1}{s^2}(\phi(0)-\phi'(0))+\frac{1}{s}\d\phi(0),\label{eq:Tphi}
\end{equation}
where $\phi'(s)$ is some matter boundary operator. The operator $\phi'(s)$
quantifies the failure of $\phi(s)$ to be a marginal operator,
or, equivalently, the failure of $\phi(s)$ to generate a conformal boundary
interaction.

\item{\bf 2)} {\it $\phi$ generates a finite boundary interaction
without renormalization.} This means that the operator
\begin{equation}\exp\left[-\int_a^b ds\, \phi(s)\right],\label{eq:2}
\end{equation}
is finite without renormalization. We assume that \eq{2} can be defined
perturbatively in powers of $\phi$.
Finiteness of \eq{2} implies that the $\phi$-$\phi$ OPE is
less singular than a simple pole:
\begin{equation}\lim_{s\to0}s\phi(s)\phi(0)=0.\label{eq:phiphi}\end{equation}
We also assume that the $\phi$-$\phi'$ and $\phi'$-$\phi'$
OPEs are less singular than a simple pole.

\item{\bf 3)} {\it $\phi$ triggers an RG flow from the reference conformal
field theory, $\mathrm{BCFT}_0$, to a target boundary conformal field theory,
$\mathrm{BCFT}^*$.} For string field
theory purposes, this means that a $\phi$ boundary interaction in correlation
functions on a very large cylinder\cite{flow1,Schnabl,Okawa} imposes
$\mathrm{BCFT}^*$ boundary conditions, while on a very small cylinder it
imposes $\mathrm{BCFT}_0$ boundary conditions. Explicitly,
\begin{equation}\lim_{L\to\infty}\left\langle\exp\left[-\int_0^L ds\,\phi(s)
\right]
\, L\circ\mathcal{O}\right\rangle^{\mathrm{BCFT}_0}_{C_L}=
\lim_{L\to\infty}\big\langle L\circ\mathcal{O}
\big\rangle^{\mathrm{BCFT}^*}_{C_L},\nonumber\end{equation}
and
\begin{equation}
\lim_{L\to0}\left\langle\exp\left[-\int_0^L ds\,\phi(s)
\right]
\, L\circ\mathcal{O}\right\rangle^{\mathrm{BCFT}_0}_{C_L}=
\lim_{L\to0}\big\langle L\circ\mathcal{O}
\big\rangle^{\mathrm{BCFT}_0}_{C_L}.\label{eq:relevant1}
\end{equation}
where $\langle\cdot\rangle^{\mathrm{BCFT}}_{C_L}$ is a
correlator on a cylinder of circumference $L$ in the
corresponding BCFT, and $L\circ\mathcal{O}$ is a scale transformation of an
arbitrary bulk operator $\mathcal{O}$ under $z\to Lz$. Scaling \eq{relevant1}
to a canonical cylinder of circumference 1,
these conditions can be equivalently stated:
\begin{equation}\lim_{u\to\infty}\left\langle\exp\left[-\int_0^1 ds\,\phi_u(s)
\right]
\,\mathcal{O}\right\rangle^{\mathrm{BCFT}_0}_{C_1} =
\big\langle \mathcal{O}\big\rangle^{\mathrm{BCFT}^*}_{C_1},\nonumber
\end{equation}
and
\begin{equation}\lim_{u\to0}\left\langle\exp\left[-\int_0^1 ds\,\phi_u(s)
\right]
\,\mathcal{O}\right\rangle^{\mathrm{BCFT}_0}_{C_1} =
\big\langle \mathcal{O}\big\rangle^{\mathrm{BCFT}_0}_{C_1},
\label{eq:relevant2}
\end{equation}
where we introduce the operator
\begin{equation}\phi_u(s) = u (u^{-1}\circ\phi(us)).\label{eq:phi_u}
\end{equation}
The parameter $u$ (equivalently $L$) can be interpreted as the RG
coupling (or time). Note that equations \eq{relevant1} and \eq{relevant2}
imply that the $\phi$ boundary interaction becomes trivial in the UV, and
therefore represents a {\it relevant} deformation of the reference
$\mathrm{BCFT}_0$. In
general, $\phi(s)$ will be a sum of different matter operators. To trigger a
flow to $\mathrm{BCFT}^*$ as described in \eq{relevant1}, the coupling
constants multiplying each component matter operator must be precisely
chosen. In the language of \cite{BMT}, fixing these couplings corresponds to
{\it tuning} the operator $\phi(s)$.

\end{description}
\noindent Conditions
{\bf 1)} and {\bf 2)} are mainly technical assumptions, and, in light of
our results, it is possible that a correct solution for the lump will not
need them. It is possible to relax {\bf 1)}
without too many further complications, but relaxing condition {\bf 2)}
would require a fundamentally different approach from the one we take here,
perhaps something analogous to the
construction of marginal solutions with singular OPEs\cite{FK,KO}.

References \cite{Ian,BMT} provide two examples of
$\phi(s)$ satisfying the above criteria. The first is the cosine
relevant deformation\cite{relevance,cosine} describing a codimension one
brane on a circle of radius $R$ greater than $\sqrt{2}$ times the
self-dual radius:
\begin{equation}\phi_u(s)= u\left[-\frac{1}{u^{1/R^2}}
:\cos\left(\frac{X(s)}{R}\right):+A(R)\right],\label{eq:cosine}\end{equation}
where $A(R)$ is a constant determined in \cite{BMT}. The restriction
$R>\sqrt{2}$ is necessary to ensure finiteness of the boundary interaction.
Unfortunately, the cosine deformation leads to an interacting worldsheet
theory, which makes it difficult to perform explicit
calculations with the solution.  A second example is the Witten
deformation\cite{Witten}, which
describes a codimension one brane along a noncompact direction:
\begin{equation}\phi_u(s)
=u\left[\frac{1}{2}:X(s)^2:+\gamma-1+\ln(2\pi u)\right].\label{eq:Witten}
\end{equation}
where $\gamma$ is the Euler-Mascheroni constant\cite{Ian}. Unlike the cosine
deformation, the Witten deformation leads to a Gaussian worldsheet theory,
and Green's functions can be computed exactly\cite{Witten}. A drawback,
however, is that the perturbative vacuum carries infinite energy
relative to the lump, and this divergence must be treated with care.

\subsection{Deformed Wedge States}
\label{subsec:alg}

The Ellwood and BMT solutions are constructed by taking star products of
the string fields
\begin{equation} K,\ \ \ \ B,\ \ \ \mathrm{and}\ \ \  c,\end{equation}
and
\begin{equation}\phi\ \ \ \mathrm{and}\ \ \ \ \phi',\end{equation}
where $K,B,c$ are defined as in \cite{simple,BMT}, and $\phi$ and $\phi'$
correspond to insertions of $\phi(s)$ and $\phi'(s)$
on the open string boundary of correlation functions on the
cylinder.\footnote{Explicitly, $\phi$ and $\phi'$ are
defined\cite{SSF1}
\begin{equation}\phi = f_\mathcal{S}^{-1}\circ\phi(\half)|I\rangle,\ \ \
\ \ \ \ \
\phi'=f_\mathcal{S}^{-1}\circ\phi'(\half)|I\rangle.\end{equation}
where $|I\rangle$ is the identity string field and
$f_\mathcal{S}^{-1}(z)=\tan\frac{\pi z}{2}$ is the inverse of the sliver
coordinate map. We use the left handed convention
for the star product\cite{simple}.} These objects satisfy
a number of identities summarized in table \ref{tab:alg}. Since $\phi(s)$ and
$\phi'(s)$ may have singular OPEs, we should be careful to avoid contact
divergences when taking open string star products.

\begin{table}[t]
\begin{center}
\begin{tabular}{|c|c|}\hline
 BRST &
$Q\phi = \d(c\phi)-\phi'\d c,\ \ \ \ \ \ \ \ Q(c\phi) = \phi'c\d c,$\\
variations & $QB=K,\ \ \ \ \ \ \ QK=0,\ \ \ \ \ \ \ Qc=c\d c.$ \\
\hline
algebraic  & $(c\phi)^2=0\ \ \ \ \ \ \
(c\phi)(\phi'c)=0$\\
identities & $Bc+cB=1,\ \ \ \ \ \ \ [K,B]=0,\ \ \ \ \ \ \ B^2=c^2=0$
\\ \hline
\end{tabular}
\end{center}
\caption{\label{tab:alg} A partial list of identities satisfied by $K,B,c$ and
$\phi$ and $\phi'$. Other relations (for example $[c,\phi]=0$)
follow easily from the definitions. The notation $\d \mathcal{O}$ indicates
the worldsheet derivative of the corresponding boundary operator
$\mathcal{O}$ in correlation
functions on the cylinder. If $\mathcal{O}$ is inserted in a segment of the
boundary with $\mathrm{BCFT}_0$ boundary conditions,
$\d\mathcal{O}=[K,\mathcal{O}]$. If $\mathcal{O}$ is inserted in a segment
with a $\phi$ boundary interaction,
$\d\mathcal{O}=[K+\phi,\mathcal{O}]$.}
\end{table}

A wedge state $\Omega^L = e^{-LK}$ is a star algebra power of the
$SL(2,\mathbb{R})$ vacuum $\Omega=|0\rangle$, and, inside correlation
functions on the cylinder, corresponds to a strip of worldsheet of
width $L>0$. A {\it deformed wedge state}
\begin{equation}\tilde{\Omega}^L = e^{-L(K+\phi)}.\end{equation}
also corresponds to strip of worldsheet of width $L>0$ on the
cylinder, but with open string boundary conditions deformed by
the $\phi(s)$ boundary interaction\cite{BMT,simple_marg}. Probing with
a Fock state,
\begin{equation}\langle\tilde{\Omega}^L,\chi\rangle = \left\langle \exp\left[
-\int_{1/2}^{L+1/2}ds\, \phi(s)\right]
f_\mathcal{S}\circ\chi(0)\right\rangle_{C_{L+1}}.\label{eq:def_Fock}
\end{equation}
Note that the $L$th star algebra power of $\tilde{\Omega}$ is directly
related to the circumference of the cylinder, and therefore to RG time.

\begin{figure}
\begin{center}
\resizebox{4in}{1.8in}{\includegraphics{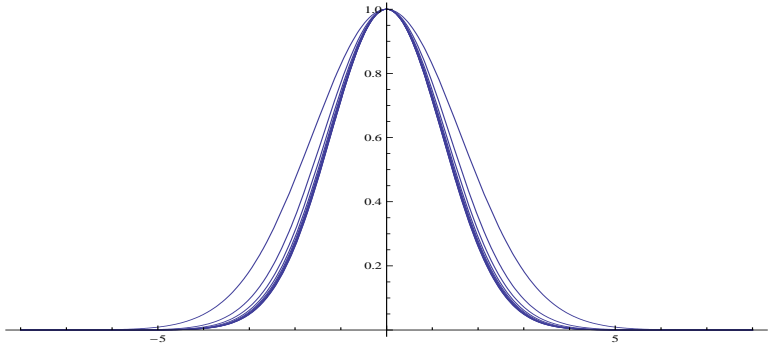}}
\end{center}
\caption{\label{fig:def_Fock} Fourier transform of the contraction
$\langle\tilde{\Omega}^L,:e^{ikX}:\rangle$ for the Witten deformation at
$u=1$,
plotted for $L=1,...,10$. The profiles are Gaussians $e^{-x^2/\Delta}$ which
become localized up to a minimum width $\Delta\approx 1.46$ for large $L$.
In this plot we have normalized the height of the Gaussians to $1$. Calculating
the absolute normalization in our approach requires evaluating the trace of
infinite dimensional matrices which are only known numerically, and
convergence is unfortunately too slow to obtain acceptable accuracy, especially
for large $L$. However, our numerics indicates that the
normalization is finite and nonzero for $L\to\infty$.}
\end{figure}

The different conformal/noncomformal boundary conditions inside \eq{def_Fock}
make the correlator difficult to compute. For the Witten deformation,
we can compute \eq{def_Fock} and similar correlators from knowledge of the
Green's function, which can be obtained as the solution of the appropriate
boundary value problem for Laplace's equation on a unit disk. We have
derived this Green's function, expressing it as a Fourier expansion whose
coefficients are products and inverses of certain infinite dimensional
matrices constructed from the Fourier modes of the boundary coupling $u$.
In practice, we must resort to numerics to compute the required matrix
inverses. With these results we can calculate the overlap of a deformed
wedge state $\tilde{\Omega}^L$ with a plane wave $:\!e^{ik X}\!:$.
Figure \ref{fig:def_Fock} shows that the overlap has a Gaussian
profile in position space, which becomes more localized, up to some minimum
uncertainty, as $L$ becomes large. As $L$ approaches infinity,
$\tilde{\Omega}^L$ approaches a constant
nonvanishing state, the {\it deformed sliver}, $\tilde{\Omega}^\infty$.

Since deformed wedge states are surface states with a nontrivial boundary
condition, one might worry that their
star products in $\mathrm{BCFT}_0$ would produce divergences analogous to
the collision of boundary condition changing operators. The
question is whether the following equation holds:
\begin{equation}\lim_{x\to0}\tilde{\Omega}^{L_1}\Omega^x\tilde{\Omega}^{L_2}
=\tilde{\Omega}^{L_1+L_2}.\end{equation}
Since the $\phi$ boundary interaction is finite without renormalization, we
expect this equation to be true in general. For the Witten deformation we
have checked it explicitly, and the limit $x\to 0$ is finite and
continuous, but non-analytic, behaving like $x^2 \ln x$.

\section{The BMT and Ellwood Solutions}
\label{sec:Ell_BMT}

Ellwood's original proposal for the lump solution\cite{Ian} is given by
writing the Schnabl gauge marginal solution\cite{Schnabl2,RZOK} as a formal
gauge transformation of the tachyon vacuum, and then relaxing the
assumption of
marginality of the matter operator inside the gauge parameter. Extending this
construction to dressed Schnabl gauges\cite{simple}
yields a class of solutions of the form\footnote{Ellwood also suggests a
further modification of the gauge parameter to impose the reality
condition\cite{Zwiebach}. The resulting solution is complicated and
we will not consider it.}
\begin{equation}\Phi_{f,g} =
f\left(c\phi-\frac{1}{K+\phi}\phi'\d c\right)
\frac{1}{1+\frac{1-fg}{K}\phi}\,Bc\, g.\label{eq:Ian_sol}
\end{equation}
We will call these {\it Ellwood solutions}. They are conjectured to be
solutions in a string field theory formulated around $\mathrm{BCFT}_0$ which
describe the endpoint of an RG flow triggered by $\phi(s)$,
$\mathrm{BCFT}^*$.  The string fields $f$ and $g$
(subject to a few conditions\footnote{We assume that $f(K)$ and $g(K)$ is a
continuous function of $K$ and $f(0)g(0)=1$. To avoid contact divergences
between $\phi$s, we must also assume $f(\infty)g(\infty)$ is finite. There
may be other conditions; see
\cite{Schnabl_lightning,exotic} for recent discussion of the algebra of wedge
states.}) can be any elements of the algebra of wedge states. Ellwood's
original solution corresponds to choosing $f$ and $g$ to be the square root
of the $SL(2,\mathbb{R})$ vacuum. The motivation behind this proposal
is a long story, discussed in depth in \cite{Ian}; essentially it amounts
to a formal argument that the solution should support the correct cohomology
of open string states. Further evidence in favor of \eq{Ian_sol} was
provided by BMT, who showed (in the special case $f=g=1$) that it has the
correct coupling to closed string states\cite{BMT}. Note that \eq{Ian_sol}
assumes the existence of an inverse for $K+\phi$. This factor contains most
of the physics of the solution, and, paradoxically, also the source of its
difficulties.

In a few special cases \eq{Ian_sol} reduces to known solutions. If $\phi$
is a marginal operator (with nonsingular OPE), $\phi'$ vanishes and
\eq{Ian_sol} reduces to the dressed Schnabl gauge marginal
solution\cite{super_marg}. Choosing $\phi=\phi'=1$ gives the dressed
Schnabl gauge tachyon vacuum solution\cite{Okawa}.

The BMT solution\cite{BMT} is a special case of \eq{Ian_sol}
with $f=g=1$:
\begin{equation}\Phi = c\phi
- B\frac{1}{K+\phi}\phi'c\d c .\label{eq:solution}
\end{equation}
This is by far the simplest solution in the class \eq{Ian_sol}. However,
the presence of the identity-like term $c\phi$ means the BMT solution is
more singular than other Ellwood solutions. However, it is not
more singular in a sense which is important for the calculation of
observables. For most of this paper (except section \ref{sec:Ellwood})
we will study the BMT solution as a prototypical example. The general Ellwood
solution can be obtained from the BMT solution by a
transformation discovered by Zeze\footnote{This transformation
first appears in a paper of Kishimoto and Michishita\cite{Zeze},
who attribute it to S. Zeze.}
\begin{equation}Z_{f,g}[\Phi]=f\Phi\frac{1}{1+A\Phi}g,\label{eq:Zeze}
\end{equation}
where $A=\frac{1-fg}{K}B$. The Zeze map is a gauge transformation only
on-shell.

The crucial ingredient in the solution is the definition of
$\frac{1}{K+\phi}$. We will follow BMT and define it (via the Schwinger
parameterization) as an integral over all deformed wedge states:
\begin{equation}\frac{1}{K+\phi}=\int_0^\infty dt\, \tilde{\Omega}^t.
\label{eq:unreg}\end{equation}
This integral converges only if the deformed sliver state
$\tilde{\Omega}^\infty$ vanishes. However, keeping in mind condition
{\bf 3)}, we would not expect the deformed sliver to vanish unless the
solution describes the tachyon vacuum. Therefore, in interesting examples,
the integral \eq{unreg} is divergent. We can regulate this divergence,
but there is another, more serious problem: Except for
tachyon vacuum solutions, the integral \eq{unreg} does not invert $K+\phi$,
\begin{equation}
(K+\phi)\left(\int_0^\infty dt\,\tilde{\Omega}^t\right)=
-\int_0^\infty dt\frac{d}{dt}\tilde{\Omega}^t=
1-\tilde{\Omega}^\infty,\label{eq:unwanted}
\end{equation}
so the definition \eq{unreg} does not accomplish its intended purpose.
This means that if we substitute \eq{unreg} in place of $\frac{1}{K+\phi}$,
the BMT and Ellwood solutions will not satisfy the equations of motion.

We can ask whether it is possible to regulate the solution so as to restore
the equations of motion. One way to do this is to replace $\phi$ with
$\phi+\eps$ for $\eps>0$ (and therefore $\phi'$ with
$\phi'+\eps$), which automatically regulates the
Schwinger integral by exponentially suppressing the $t=\infty$
boundary of moduli space. The expression
\begin{equation}\Psi_\mathrm{tv}(\eps)=c(\phi+\eps)-B\frac{1}{K+\phi+\eps}
(\phi'+\eps)c\d c\label{eq:tv}\end{equation}
is a solution to the equations of motion for all $\eps>0$,
since $\phi+\eps$ is simply another choice of $\phi$. Unfortunately, this
regularization drastically alters the physical interpretation of the
solution: For all $\eps>0$, this is a (non universal)
solution for the tachyon vacuum. To see this, note that the
boundary interaction of $\phi(s)+\eps$ on a very large cylinder vanishes
(assuming $\phi(s)$ is already tuned), thanks to the divergent integration
of $\eps$ along the boundary.

Paradoxically, the unwanted sliver term in \eq{unwanted} which threatens the
equations of motion is physically necessary. If it were possible
to invert $K+\phi$, we could trivialize the cohomology around the
BMT solution with the homotopy operator\cite{BMT}
\begin{equation}\frac{B}{K+\phi}.\end{equation}
Furthermore, the same mechanism which produces the unwanted boundary
term also
accounts for the correct IR coupling to closed string states\cite{BMT}.
Therefore in this setup there is a basic tension between the equations of
motion and the desire to reproduce the physics of a nontrivial boundary
conformal field theory in the infrared. This observation is already
enough to exclude the BMT and Ellwood solutions as viable descriptions of
the lump.
Undoubtedly, another class of solutions, perhaps even closely related, can
resolve this problem, but our goal in this
paper is more narrow: We will show that, in spite of the failure of the
equations of motion, the BMT and Ellwood solutions reproduce essentially
all of the expected physics of the lump, when properly interpreted. The
significance of this observation is currently unclear to us, but
it seems potentially important.

Given that $\frac{1}{K+\phi}$ is divergent in the Fock space, one might
worry that the BMT and Ellwood solutions are also divergent. Actually,
this is not the case, because in \eq{Ian_sol} $\frac{1}{K+\phi}$ always
appears multiplied by $B$, which annihilates the linear
divergence proportional to the deformed sliver state ($B\tilde{\Omega}^t$
vanishes as $1/t^3$ for large $t$). As we will see, this
cancellation is not enough to restore the equations of motion.
In fact, it is in a sense accidental. There are other solutions
which are in principle equivalent to Ellwood and BMT but actually diverge in
the Fock space. Consider for example the solution
\begin{eqnarray}\hat{\Phi}
=c\phi -
\frac{1}{2}B\frac{1}{K+\phi}\phi'c\d c
-\frac{1}{2}\phi'c\d c\frac{1}{K+\phi}B+\frac{1}{4}B\frac{1}{K+\phi}\phi'
\d c\frac{1}{K+\phi}\phi'\d c\frac{1}{K+\phi},\nonumber\\
\end{eqnarray}
which, unlike \eq{Ian_sol}, satisfies the string field reality condition.
Regulating $\frac{1}{K+\phi}$ with a cutoff, one can easily show that this
expression has the correct coupling to closed string states. Nevertheless,
unless $\phi$ triggers an RG flow to the tachyon vacuum, this solution
diverges in the Fock space since the last term has too many powers of
$\frac{1}{K+\phi}$. At present we do not know of any solution describing a
nontrivial $\mathrm{BCFT}^*$ which is both real and finite.

\section{BMT solution}
\label{sec:BMT}

\subsection{Regularization and Closed String Overlap}
\label{subsec:BMT_reg}
The BMT and Ellwood solutions \eq{solution} and \eq{Ian_sol}
are not well-defined string fields as they stand (except when they describe
the tachyon vacuum), and to give them meaning we must apply some
regularization. Implicitly the computations of \cite{BMT} regulate
by imposing a hard cutoff for the Schwinger integral \eq{unreg}. However,
a finite cutoff is cumbersome for most calculations. Instead, we will consider
the BMT solution as the $\eps\to 0$ limit of the string field
\begin{equation}\Phi(\eps)=c\phi - B\frac{1}{K+\phi+\eps}\phi'c\d c,
\label{eq:solution2}\end{equation}
where
\begin{equation}\frac{1}{K+\phi+\eps}=\int_0^\infty dt\,e^{-\eps t}
\tilde{\Omega}^t.\label{eq:soft}\end{equation}
Here $\eps$ regulates the Schwinger integral by exponentially
suppressing the $t=\infty$ boundary of moduli space. Therefore
\eq{solution2} is a finite, well-defined string field for all $\eps>0$.
We will discuss the regularization of Ellwood solutions in section
\ref{sec:Ellwood}.

The nice thing about the regularization \eq{solution2} is that
$\Phi(\eps)$
can be neatly separated into two terms: the tachyon
vacuum solution $\Psi_\mathrm{tv}(\eps)$ in \eq{tv}, and a term $\Delta(\eps)$
which ``builds'' the lump on top of the tachyon vacuum:
\begin{equation}\Phi(\eps) = \Psi_\mathrm{tv}(\eps)+\Delta(\eps),
\label{eq:PhiPsi}\end{equation}
where
\begin{equation}\Delta(\eps) = -\eps c +B\frac{\eps}{K+\phi+\eps}c\d c.
\label{eq:Delta}\end{equation}
In the $\eps\to 0$ limit $\Delta(\eps)$ becomes a sliver-like state, but
vanishes in the level expansion\footnote{The factor
$\frac{\eps}{K+\phi+\eps}$ approaches the
deformed sliver as $\eps\to 0$. Since $B$ annihilates the deformed sliver
when contracted with Fock states, $\Delta(\eps)$ also vanishes.}.
Nevertheless it has a crucial effect on the calculation of observables.
In this sense it is a kind of ``phantom term.''
Note the differing roles $\eps$ plays in \eq{PhiPsi}. In
$\Phi(\eps)$ it plays the role of a regulating
parameter, but in $\Psi_\mathrm{tv}(\eps)$ it is a
gauge parameter labeling a class of equivalent solutions for the
tachyon vacuum.

Equation \eq{PhiPsi} is very useful for calculating observables, since
it allows us to cleanly separate the ``trivial'' contribution from the tachyon
vacuum from the physically interesting contribution of the lump. To illustrate
this point, let us present an alternative computation of the closed string
overlap, which was already computed in \cite{BMT}. Recall from
\cite{Ellwood,BMT} that the closed string overlap of the BMT solution should
satisfy
\begin{equation}
\lim_{\eps\to 0}\Tr[\V\Phi(\eps)]=-\mathcal{A}_0(\mathcal{V})
+\mathcal{A}^*(\mathcal{V}),
\label{eq:overlap}\end{equation}
where $\mathcal{A}_0(\mathcal{V})$ is the disk amplitude in $\mathrm{BCFT}_0$
with one on-shell closed string insertion
$\mathcal{V}=c\tilde{c}\mathcal{V}^\mathrm{m}$, $\mathcal{A}^*(\mathcal{V})$
is the same quantity in $\mathrm{BCFT}^*$,
and $\Tr[\mathcal{V}\cdot]$ is the
1-string vertex with a midpoint insertion of $\mathcal{V}$.
Plugging the regularized BMT solution into the right hand side of
\eq{overlap}
we find two terms:
\begin{equation}\Tr[\V\Phi(\eps)]
=\Tr[\V\Psi_\mathrm{tv}(\eps)]+\Tr[\V\Delta(\eps)].
\end{equation}
The first term is the closed string overlap of a tachyon vacuum
solution. Since the disk tadpole amplitude around the tachyon vacuum
vanishes, the first term only contributes minus the disk amplitude around
$\mathrm{BCFT}_0$. Therefore,
\begin{equation}\Tr[\V\Phi(\eps)]
=-\mathcal{A}_0(\mathcal{V})+\Tr[\V\Delta(\eps)].
\end{equation}
Without any calculation we are already half-way done. The second term from
$\Delta(\eps)$ must exclusively account for
the nontrivial coupling between closed strings and $\mathrm{BCFT}^*$.
Modulo ghost factors, we can already see that this is essentially
guaranteed since $\Delta(\eps)$ is proportional to the deformed
sliver state in the $\eps\to 0$ limit, and when we take the trace the
boundary conditions are driven to the IR, producing the disk tadpole
amplitude in $\mathrm{BCFT}^*$.  Let us see this explicitly:
\begin{eqnarray}\Tr[\V\Delta(\eps)]
=\eps\int_0^\infty dt\,e^{-\eps t} \Tr[\V\tilde{\Omega}^t Bc\d c].
\label{eq:gh_D}
\end{eqnarray}
Simplifying the ghost correlator\footnote{The $-\eps c$ term does not
contribute because of the negative conformal dimension of $c$\cite{simple}.
Note
\begin{equation}\tilde{\Omega}^t Bc\d c = -\frac{1}{t}\tilde{\Omega}^t c
-\frac{1}{t}\half\mathcal{B}^-(\tilde{\Omega}^tc\d c),\end{equation}
where $\mathcal{B}^-$ is defined in \cite{simple}. Using $\mathcal{B}^-$
invariance of the vertex, the second term does not contribute to the
trace, leading to \eq{d1}.} and substituting $\alpha=\eps t$ this becomes
\begin{equation}\Tr[\V\Delta(\eps)]
=\int_0^\infty d\alpha\,e^{-\alpha}\left(-\,
\frac{\eps}{\alpha}\Tr[\V
\tilde{\Omega}^{\alpha/\eps}c]\right).
\label{eq:d1}\end{equation}
The factor in the integrand can be written as a correlation function on the
cylinder:
\begin{equation}-\,\frac{\eps}{\alpha}\Tr[\V
\tilde{\Omega}^{\alpha/\eps}c]=-\left\langle
\exp\left[\int_0^{\alpha/\eps}ds\,\phi(s)\right]
\left(\frac{\alpha}{\eps}\right)
\circ\Big[\mathcal{V}(i\infty)c(0)\Big]\right\rangle_{C_{\alpha/\eps}}.
\end{equation}
In the $\eps\to 0$ limit the cylinder becomes very large, and the boundary
conditions flow to the IR as described in condition {\bf 3)}. Thus,
\begin{equation}\lim_{\eps\to 0}\left(-\,\frac{\eps}{\alpha}\Tr[
\V
\tilde{\Omega}^{\alpha/\eps}c]\right)=-\langle
\mathcal{V}(i\infty)c(0)\rangle_{C_1}^{\mathrm{BCFT}^*}.\end{equation}
The right hand side is precisely the amplitude for a single closed string to
be emitted from the background
$\mathrm{BCFT}^*$, as defined in the conventions of \cite{Ellwood}.
Integrating over $\alpha$ gives
\begin{equation}\lim_{\eps\to0}\Tr[\V\Delta(\eps)]
=\mathcal{A}^*(\mathcal{V}).\end{equation}
In total
\begin{equation}
\lim_{\eps\to0}\Tr[\V\Phi(\eps)]
=-\mathcal{A}_0(\mathcal{V})+\mathcal{A}^*(\mathcal{V}),
\end{equation}
as expected.  Unlike the closed string overlap, the energy is a nonlinear
function of the string field, and its separation into a tachyon vacuum
and lump contribution is less obvious. We will see how this happens in section
\ref{subsec:energy}.

\subsection{Anomaly in Equations of Motion}
\label{subsec:anomaly}

The regularization \eq{solution2} does not produce a
solution to the equations of motion, since, despite formal appearances,
$\frac{1}{K+\phi+\eps}$ does not define an inverse for $K+\phi$ in the
$\eps\to 0$ limit. We can check this:
\begin{equation}(K+\phi)\frac{1}{K+\phi+\eps} = 1-\frac{\eps}{K+\phi+\eps}.
\end{equation}
The second term is not zero because the vanishing of $\eps$ is
compensated by a linear divergence in $\frac{1}{K+\phi+\eps}$. To see what
happens explicitly, plug in the Schwinger
integral for $\frac{1}{K+\phi+\eps}$ and make a substitution
$\alpha=\eps t$:
\begin{equation}\frac{\eps}{K+\phi+\eps} = \int_0^\infty d\alpha\,
 e^{-\alpha}
\tilde{\Omega}^{\alpha/\eps}.\label{eq:subtrick}\end{equation}
Note that the overall factor of $\eps$ has disappeared.
Taking $\eps\to 0$, the deformed wedge state in the integrand becomes the
deformed sliver, independent of $\alpha$. Integration over $\alpha$ then
produces a factor of 1 and we find
\begin{equation}\lim_{\eps\to 0}(K+\phi)\frac{1}{K+\phi+\eps}
= 1-\tilde{\Omega}^\infty,\end{equation}
exactly as we found in equation \eq{unwanted}.

It is important to calculate precisely how the equations of motion
fail. For this purpose, write the regularized BMT
solution in the form
\begin{equation}\Phi(\eps)=c\phi - A(\eps)\phi'c\d c,\ \ \ \ \ \ \ \ \
A(\eps)=\frac{B}{K+\phi+\eps},\end{equation}
where $A(\eps)$ is the homotopy operator which trivializes the cohomology
around the tachyon vacuum solution $\Psi_\mathrm{tv}(\eps)$ in \eq{tv}:
\begin{equation} Q_{\Psi_\mathrm{tv}(\eps)}A(\eps)=1.
\label{eq:hom_tv}\end{equation}
Now note from table \ref{tab:alg} that the kinetic term of the equations
of motion can be written
\begin{equation}Q\Phi(\eps) = (1-QA(\eps))\phi'c\d c,\label{eq:quad1}
\end{equation}
while the quadratic term can be written
\begin{equation}\Phi(\eps)^2 = -[\Phi(\eps),A(\eps)]\phi'c\d c.
\label{eq:inter1}\end{equation}
The commutator of the right hand side is there for free,
because the product of $\Phi(\eps)$ with $\phi'c\d c$ vanishes upon the
collision of $c$s.
Adding \eq{quad1} and \eq{inter1} together, we find
\begin{equation}Q\Phi(\eps)+\Phi(\eps)^2 = \Big[1-Q_{\Phi(\eps)}A(\eps)\Big]
\phi'c\d c.\label{eq:an1}
\end{equation}
Therefore, the validity of the the equations of motion is directly
related to question of whether the BMT solution supports open string
states. If it does support open string states, then
$Q_{\Phi(\eps)}A(\eps)$ cannot be $1$ since otherwise $A(\eps)$ would
trivialize the cohomology. But then \eq{an1} implies the equations of
motion cannot be satisfied.

Let us complete the computation of \eq{an1}:
\begin{eqnarray}Q_{\Phi(\eps)}A(\eps)\lineup
= Q_{\Psi_\mathrm{tv}(\eps)}A(\eps)+[\Delta(\eps),A(\eps)],\nonumber\\
\lineup =1+[\Delta(\eps),A(\eps)].\end{eqnarray}
The commutator is explicitly,
\begin{equation}[\Delta(\eps),A(\eps)] = -cB\frac{\eps}{K+\phi+\eps}
-\frac{\eps}{K+\phi+\eps}Bc +\frac{\eps}{K+\phi+\eps}\d c
\frac{B}{K+\phi+\eps}.
\end{equation}
Now write $\d c = [K+\phi+\eps,c]$. This allows us to
cancel one of the inverse factors of $K+\phi+\eps$ in the third term, and
adding up what remains gives simply
\begin{equation}[\Delta(\eps),A(\eps)] =
-\frac{\eps}{K+\phi+\eps}.\end{equation}
Therefore,
\begin{equation}Q_{\Phi(\eps)}A(\eps)
= 1-\frac{\eps}{K+\phi+\eps},\end{equation}
and
\begin{eqnarray}Q\Phi(\eps)+\Phi(\eps)^2\lineup
=\frac{\eps}{K+\phi+\eps}\phi'c\d c,\nonumber\\
\lineup \equiv \Gamma(\eps),\label{eq:anomaly}
\end{eqnarray}
where $\Gamma(\eps)$ is the anomaly in the equations of motion.

For $\eps>0$, the anomaly is (of course) nonzero, since $\Phi(\eps)$ is not
intended to be a solution for nonzero $\eps$. In the limit $\eps\to 0$
we find
\begin{equation}\Gamma(0)=
\tilde{\Omega}^\infty \phi'c\d c\label{eq:gamma0}.\end{equation}
In \cite{BMT} it was implicitly assumed that this state vanishes. The
intuition behind this expectation is that the deformed sliver tends to
drive the boundary conditions to the IR, where $\phi'$ vanishes by
conformal invariance. Nevertheless, \eq{gamma0} is a
nonvanishing state. The
reason is because $\phi'$ appears at the outer edge of the deformed sliver,
where test states do not see the boundary conditions as conformal. More
formally, since $\phi'$ is on the edge it can have contractions with
nearby operators which do not vanish in the sliver limit, even though
the 1-point function of $\phi'$ does vanish.

\begin{figure}
\begin{center}
\resizebox{3.5in}{1.5in}{\includegraphics{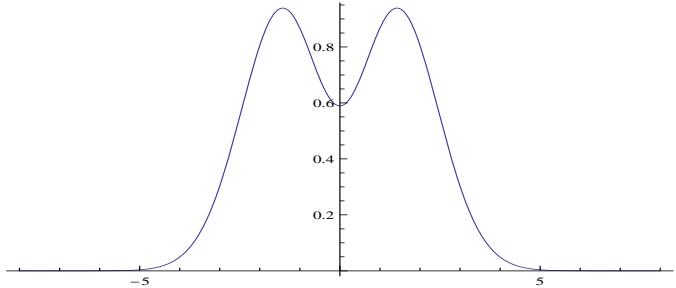}}
\end{center}
\caption{\label{fig:an} Fourier transform of the contraction
$\langle \Gamma(0), c\!\!:\!\!e^{ikX}\!\!:\rangle$ assuming the Witten deformation at
$u=1$. We were
not able to accurately determine the normalization, but we believe it is
finite and nonzero. We are not certain whether the double peak profile
has a useful interpretation, though it is interesting to note that
the equations of motion appear to be satisfied far away from the center of
the lump. Perhaps not coincidentally, far away from the position of the brane
we expect the solution to approximate the tachyon vacuum.}
\end{figure}

\begin{figure}
\begin{center}
\resizebox{3in}{1.5in}{\includegraphics{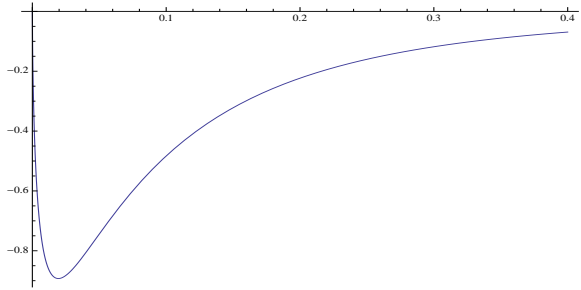}}
\end{center}
\caption{\label{fig:lump1} Anomaly in the equations of motion contracted
with the state \eq{T}, plotted as a function of the boundary coupling $u$
of the Witten deformation \eq{Witten} for $0<u<.4$.}
\end{figure}

For the Witten deformation, we have calculated the overlap of $\Gamma(0)$
with a plane wave, giving the position space profile shown
in figure \ref{fig:an}. However, it is possible to see that the anomaly
is nonvanishing with a simpler calculation.
For the Witten deformation, consider a test state
 \begin{equation}\chi
 = \tilde{\Omega}^{1/2}(:\!X^2\!:\!\d^2 c)\tilde{\Omega}^{1/2},
\label{eq:T}\end{equation}
where the strips on either side of the insertion are deformed. Contracting
this with $\Gamma(0)$ gives a correlator whose entire boundary shares the
same nonconformal boundary condition, and we can compute
the correlator using formulas originally written down by Witten\cite{Witten},
reviewed in appendix \ref{app:Witten}. The result is
\begin{equation}\langle \Gamma(0),\chi\rangle
=  -16 u \Big[\cos(2\pi u)\ci(2\pi u)+\sin(2\pi u)\si(2\pi u)
\Big]^2,\end{equation}
where $u$ is the boundary coupling of the Witten deformation
and $\ci$ and $\si$ are the sine and cosine integrals
\begin{equation}\ci(x)=-\int_x^\infty ds\frac{\cos s}{s}\ \ \ \ \ \ \ \
\si(x)=-\int_x^\infty ds\frac{\sin s}{s}.\label{eq:sici}\end{equation}
We plot this as a function $u$ in figure \ref{fig:lump1}.
Note that the overlap vanishes for very large $u$:
\begin{equation}\langle \Gamma(0),\chi\rangle \sim -\frac{1}{\pi^4 u^3}
\ \ \ (\mathrm{large}\ u).
\end{equation}
This suggests that there is a sense in which the equations of motion are
satisfied when formulating the solution directly at the infrared fixed
point of the RG flow. We will say more about this when
discussing the cohomology.

\subsection{Energy}
\label{subsec:energy}

Leaving the equations of motion to the side for a moment, let us proceed
to calculate the energy of the BMT solution. For static backgrounds,
we can calculate the energy by evaluating the action, and the
result should be\cite{BMT}
\begin{equation}E = -S =
\frac{1}{2\pi^2}(-g(0)+g(\infty)),\end{equation}
where $g(0)$ is the norm of the $SL(2,\mathbb{R})$ vacuum in
$\mathrm{BCFT}_0$ and $g(\infty)$ is the norm of the $SL(2,\mathbb{R})$
vacuum in $\mathrm{BCFT}^*$. Note that $g(0)$ and $g(\infty)$ are related
via RG flow of the disk partition function
\begin{equation}g(L) = \Tr[\tilde{\Omega}^L]^m,\end{equation}
where $\Tr[\cdot]^m$ is the trace in the matter component of
$\mathrm{BCFT}_0$.\footnote{Implicitly, we assume that matter correlators
include a trivial ghost factor to ensure vanishing central charge, and
likewise ghost correlators contain a trivial matter factor.}

Since the equations of motion are not satisfied, there is no reason to
expect to find
the correct energy by computing the kinetic or cubic terms alone.
Therefore, we will calculate the full gauge invariant action
\begin{equation}S=\Tr\left[-\frac{1}{2}\Phi(\eps)Q\Phi(\eps)-\frac{1}{3}
\Phi(\eps)^3\right].\end{equation}
in the limit $\eps\to 0$.

Keeping track of the anomaly in the equations of motion, we can write the
action
\begin{equation}S = \Tr\left[\frac{1}{6} \Phi(\eps)^3
-\frac{1}{2}\Phi(\eps)\Gamma(\eps)\right].\end{equation}
Our strategy is to use the decomposition
\begin{equation}\Phi(\eps)=\Psi_\mathrm{tv}(\eps)+\Delta(\eps)
\label{eq:PhiPsi2}\end{equation}
to separate the action into a contribution from the tachyon vacuum and
a contribution from the lump. We start by substituting this decomposition
into the $\Phi(\eps)^3$ term:
\begin{eqnarray}S\lineup = \Tr\left[\frac{1}{6}(\Psi_{\mathrm{tv}}(\eps)+
\Delta(\eps))^3
-\frac{1}{2}\Phi(\eps)\Gamma(\eps)\right], \nonumber\\
\lineup=\Tr\left[\frac{1}{6}\Psi_{\mathrm{tv}}(\eps)^3
+\frac{1}{2}\Delta(\eps)\Psi_{\mathrm{tv}}(\eps)^2+\frac{1}{2}\Delta(\eps)^2\Psi_{\mathrm{tv}}(\eps)
+\frac{1}{6}\Delta(\eps)^3
-\frac{1}{2}\Phi(\eps)\Gamma(\eps)\right].
\end{eqnarray}
Now write the last four terms back in terms of $\Phi(\eps)$, keeping the
$\Psi_\mathrm{tv}(\eps)^3$ term:
\begin{equation}S=\Tr\left[\frac{1}{6}\Psi_{\mathrm{tv}}(\eps)^3\right]
+\Tr\left[\frac{1}{2}\Delta(\eps)\Phi(\eps)^2-\frac{1}{2}\Delta(\eps)^2
\Phi(\eps)+\frac{1}{6}\Delta(\eps)^3
-\frac{1}{2}\Phi(\eps)\Gamma(\eps)\right]. \label{eq:step1}\end{equation}
The first term is the gauge invariant action evaluated on a solution for
the tachyon vacuum. Without any calculation, we know what this quantity
is: it is the vacuum energy of the reference $\mathrm{BCFT}_0$:
\begin{equation}\frac{1}{6}\Tr[\Psi_{\mathrm{tv}}(\eps)^3]
=\frac{1}{2\pi^2}g(0).
\label{eq:vac_energy}\end{equation}
We will take this equation as given. We do not have
a general proof except to note that $\Psi_\mathrm{tv}(\eps)$ is
related to other solutions for the tachyon vacuum, whose energy has been
computed analytically, by gauge transformations which
we believe are nonsingular and can be continuously deformed to the
identity.\footnote{For example, the ``simple'' tachyon
vacuum of \cite{simple} can be expressed as a limit
$\lim_{\eps\to\infty}\eps^{\frac{1}{2}\mathcal{L}^-}
\Psi_\mathrm{tv}(\eps)$, which is a combination of a reparameterization
and a shift in the gauge parameter $\eps$. Note that the gauge transformation
needed to relate these two solutions is not unique.}
For the Witten deformation
we are able to verify \eq{vac_energy} numerically, as discussed in appendix
\ref{app:vac_energy}.

Assuming \eq{vac_energy}, all we have to do is show that the four remaining
terms in \eq{step1} conspire to give the energy of the lump. To simplify,
we use the identity
\begin{equation}\Delta(\eps)\Phi(\eps)
=\Gamma(\eps).\label{eq:dpg}\end{equation}
This equality is easy to verify using the relations of table \ref{tab:alg},
though we don't have an interpretation of why it holds.\footnote{Note a
curious thing: The anomaly $\Gamma(0)$ is a nonzero state, even
though $\Delta(\eps)$ vanishes in the level expansion when $\eps\to 0$. This
suggests that the BMT solution is in a sense ``infinite,'' even though it
is finite in the level expansion.} There does not seem to be an analogous
relation for Ellwood solutions. At any rate, plugging \eq{dpg} into
\eq{step1}, the term with the anomaly contracted with the solution cancels,
and we are left with
\begin{equation}S = \frac{1}{2\pi^2}g(0) +
\Tr\left[-\frac{1}{2}\Delta(\eps)\Gamma(\eps)
+\frac{1}{6}\Delta(\eps)^3\right].\label{eq:step2}
\end{equation}
On the right hand side, both terms in the trace are sliver-like, and so
they drive the boundary conditions to the IR. The term with the anomaly
however has an insertion of $\phi'$, which tends to kill the correlator
as the boundary conditions become conformal. Therefore we can anticipate
that only the $\Delta(\eps)^3$ term contributes to the energy of the lump.
This is exactly as we would expect if $\Delta(\eps)$ were a genuine
solution of the equations of motion for the lump expanded around the
tachyon vacuum. However, as we know from previous discussion,
it is not.

We now compute the right hand side of \eq{step2} explicitly. Start with
the term
\begin{eqnarray}\Tr[ \Delta(\eps)\Gamma(\eps)]\lineup =
\Tr\left[\frac{\eps}{K+\phi+\eps}Bc\d c\frac{\eps}{K+\phi+\eps}\phi'c\d c
\right].
\end{eqnarray}
Expanding out the Schwinger integrals, making a change of variables,
and separating the trace into matter and ghost components gives the
expression
\begin{equation}\Tr[ \Delta(\eps)\Gamma(\eps)]
=\frac{1}{\eps}\int_0^\infty d\alpha\,\alpha^2 e^{-\alpha}
\Tr[\tilde{\Omega}^{\alpha/\eps}\phi']^m \int_0^1 dq
\Tr[\Omega^{1-q} Bc\d c \Omega^{q} c\d c]^{gh}.\end{equation}
Evaluating the ghost integral,
\begin{equation}
\Tr[ \Delta(\eps)\Gamma(\eps)]
=-\frac{1}{2\pi^2}\frac{1}{\eps}\int_0^\infty d\alpha\,\alpha^2 e^{-\alpha}
\Tr[\tilde{\Omega}^{\alpha/\eps}\phi']^m.\end{equation}
To get rid of the $\phi'$ use the relation\footnote{This can be shown
using invariance of the
vertex under reparameterizations by $\mathcal{L}^-$\cite{simple}.
Noting that $\mathcal{L}^-$ is a derivation and that
$\half\mathcal{L}^-\phi = \phi-\phi'$ and $\half\mathcal{L}^-K=K$ we can
show
\begin{equation}\half\mathcal{L}^-\tilde{\Omega}^L = \int_0^L ds\,
\tilde{\Omega}^{L-s}\phi'\tilde{\Omega}^s+L\frac{d}{dL}\tilde{\Omega}^L
\label{eq:446}
\end{equation} Taking the trace of both sides, $\mathcal{L}^-$ kills the
vertex and we are left with \eq{phipdL}.}
\begin{equation}\Tr[\tilde{\Omega}^L\phi']^m =
-\frac{d}{dL}g(L).\label{eq:phipdL}\end{equation}
Note that this implies that the 1-point function of $\phi'$ on a very large
cylinder vanishes, since the disk partition function approaches a constant
in the infrared (the IR norm of the $SL(2,\mathbb{R})$ vacuum). This is the
sense in which the anomaly was naively expected to vanish. Plugging
\eq{phipdL} into \eq{446} we find
\begin{equation}
\Tr[ \Delta(\eps)\Gamma(\eps)]
=-\frac{1}{2\pi^2}\int_0^\infty d\alpha\,\alpha e^{-\alpha}
\left.\left[L\frac{d}{dL}g(L)\right]
\right|_{L=\alpha/\eps}.\end{equation}
Since $g(L)$ approaches a constant its derivative
must vanish faster than $1/L$. Therefore
\begin{equation}\lim_{\eps\to 0}\Tr[\Delta(\eps)\Gamma(\eps)]=0.
\label{eq:SEOM}\end{equation}
From the perspective of the gauge invariant action, the anomaly
in the equations of motion vanishes.

Finally, compute the term
\begin{equation}\frac{1}{6}\Tr[\Delta(\eps)^3]
=\frac{1}{6}\Tr\left[\left(\frac{\eps}{K+\phi+\eps}Bc\d c\right)^3\right].
\end{equation}
Expanding the Schwinger integrals, and making a change of variables, and
separating into matter/ghost components,
\begin{equation}\frac{1}{6}\Tr[\Delta(\eps)^3]=
\frac{1}{6}\int_0^\infty d\alpha\, \alpha^2 e^{-\alpha}
g\left(\fraction{\alpha}{\eps}\right)
\int_0^1 dq\int_0^qdr \Tr[\Omega^{1-q}Bc\d c\Omega^{q-r}B c\d c
\Omega^{r}Bc\d c]^{gh}.
\end{equation}
Evaluating the ghost integral
\begin{equation}
\int_0^1 dq\int_0^qdr \Tr[\Omega^{1-q}Bc\d c\Omega^{q-r}B c\d c
\Omega^{r}Bc\d c]^{gh}=-\frac{3}{2\pi^2}
\end{equation}
gives
\begin{equation}\frac{1}{6}\Tr[\Delta(\eps)^3]=
-\frac{1}{4\pi^2}\int_0^\infty d\alpha\, \alpha^2 e^{-\alpha}
g\left(\fraction{\alpha}{\eps}\right).
\end{equation}
In the $\eps\to 0$ limit the disk partition function becomes the norm of the
$SL(2,\mathbb{R})$ vacuum in the infrared, and the integration over $\alpha$
gives a factor of $2!$. Thus
\begin{equation}\lim_{\eps\to 0}\frac{1}{6}\Tr[\Delta(\eps)^3]=
-\frac{1}{2\pi^2}g(\infty).\label{eq:D3}
\end{equation}
and the total energy is
\begin{equation}E = -S =
\frac{1}{2\pi^2}(-g(0)+g(\infty)),\label{eq:energy}\end{equation}
as expected.

We may ask how this result depends on our choice of
regularization. It depends to some degree, since the tachyon
vacuum solution \eq{tv} can be viewed as a regularization of the BMT
solution. That being said, we believe that the lump energy works for
a large class of regularizations of the BMT solution. For example, with an
assumption\footnote{The needed assumption is that $\Tr[\Phi^3]$ is
an absolutely convergent integral over correlation functions on the
cylinder. This is true for the Witten deformation.},
the energy works for
any regularization which represents $\frac{1}{K+\phi}$ as a limit of states
in the deformed wedge algebra while leaving the rest of the solution
unchanged. This includes, for example, regulating the solution with a hard
cutoff for the upper limit of the Schwinger integral \eq{unreg}.
Nevertheless, the regularization we have used produces highly nontrivial
simplifications which may have a deeper explanation.

\subsection{Equations of Motion Contracted with the Solution}
\label{subsec:cubic}

It is interesting to ask whether we would have obtained the correct energy
calculating only the cubic or kinetic terms of the action. The answer to this
question depends on whether the anomaly contracted with the solution,
\begin{equation}\lim_{\eps\to 0}\Tr[\Phi(\eps)\Gamma(\eps)],\label{eq:an_sol}
\end{equation}
vanishes. We see no reason why this quantity should vanish in general, but
its explicit computation depends on the choice of relevant deformation and,
since it is not gauge invariant, the particular form of the BMT solution.
Here we compute \eq{an_sol} using the Witten deformation.

Plugging in the anomaly and BMT solution gives
\begin{eqnarray}\Tr[\Phi(\eps)\Gamma(\eps)]\lineup =
-\Tr\left[\frac{B}{K+\phi+\eps}\phi'c\d c\frac{\eps}{K+\phi+\eps}\phi'c\d c
\right].
\end{eqnarray}
Now on the right hand side we expand the Schwinger integrals and separate
them into an integration
over the total width of the cylinder and an integration over
the relative positions of the $\phi'c\d c$ insertions within a cylinder
of fixed width. In the $\eps\to 0$ limit only a cylinder with infinite
circumference contributes to the integral, and we are left only with the
integral over the separation of the $\phi'c\d c$ insertions:
\begin{equation}\lim_{\eps\to0}\Tr[\Phi(\eps)\Gamma(\eps)]
=-\lim_{L\to\infty}\left(L\int_0^1 dq\,\Tr[B\tilde{\Omega}^{Lq}\,\phi'c\d c
\,\tilde{\Omega}^{L(1-q)}\,\phi'c\d c]\right).
\end{equation}
The integration variable $q$ is the ratio between the circumference
of the cylinder and the separation between the $\phi'c\d c$ insertions.
To go further we need to evaluate the matter correlator, which requires us
to specialize to the Witten deformation. The 2-point function of $\phi'$ in
the presence of the Witten boundary interaction can be easily computed
from \eq{2-point} in appendix \ref{app:Witten}. Dropping the contribution
from the square of the 1-point function of $\phi'$, which vanishes as
$1/L^4$, and computing the ghost correlator leaves the
integral\footnote{With the appropriate substitution we set the coupling
parameter $u$ of the Witten deformation to unity.}
\begin{equation}
\lim_{\eps\to0}\Tr[\Phi(\eps)\Gamma(\eps)]
=-\frac{g(\infty)}{2\pi^2}\lim_{L\to\infty}\left(L^2\int_0^1 dq\,
\Big[\pi(1-q)\sin 2\pi q+\cos2\pi q-1\Big]G(Lq,L)^2 \right),
\end{equation}
where $G(s,L)$ is the boundary Green's function for the Witten
deformation of two $X$ insertions separated by a distance $s$ on a
cylinder of circumference $L$:
\begin{equation}\left\langle e^{-\int_0^L dt\,\phi_u(t)}X(s)X(0)
\right\rangle_{C_L}^{\mathrm{BCFT}_0}=g(L)G(s,L).
\end{equation}
Its explicit form is given in
\eq{G_boundary}. For generic $q$ the Green's function in the
integrand vanishes for large $L$, because the boundary conditions on a very
large cylinder are effectively Dirichlet which forces $X=0$.
The integral therefore can only receive contribution from singular
contractions between two $X$s in the vicinity of $q=0$ and $q=1$.
Furthermore, at $q=1$ the singular contraction between the $X$s is suppressed
by a vanishing contraction between the pair of $c\d c$s. Therefore, in the
large $L$ limit the integral only receives contribution from the vicinity
of $q=0$. To clearly extract this contribution, we make a substitution of
variables $x=2\pi L q$:
\begin{equation}
\lim_{\eps\to0}\Tr[\Phi(\eps)\Gamma(\eps)]
=-\frac{g(\infty)}{4\pi^2}\lim_{L\to\infty}\left(L\int_0^{2\pi L} dx\,
\frac{1}{\pi}\left[\pi\left(1-\frac{x}{2\pi L}\right)\sin
\frac{x}{L}+\cos\frac{x}{L}-1\right]G\left(\frac{x}{2\pi},L\right)^2 \right).
\end{equation}
Making a series expansion of the trigonometric factor in the integrand, the
integral further simplifies
\begin{equation}
\lim_{\eps\to0}\Tr[\Phi(\eps)\Gamma(\eps)]
=-\frac{g(\infty)}{4\pi^2}\int_0^\infty dx\, x\,
\lim_{L\to\infty}G\left(\frac{x}{2\pi},L\right)^2.
\end{equation}
Thus we need to know the 2-point function of $X$s with fixed separation
on a very large cylinder. This limit is given in \eq{fixed_s}. Thus the
equations of motion contracted with the solution is
\begin{equation}\lim_{\eps\to0}\Tr[\Phi(\eps)\Gamma(\eps)]
=-\frac{g(\infty)}{\pi^2}w,\label{eq:PhiG}\end{equation}
where $w$ is the value of the integral
\begin{equation}
w=\int_0^\infty dx\, x(\cos (x)\,\ci(x)+\sin (x)\si(x))^2=0.36685.
\end{equation}
This means that if we had evaluated the cubic term in the action expecting
to find the correct energy, instead we would have found
\begin{eqnarray}-\lim_{\eps\to0}\frac{1}{6}\Tr[\Phi(\eps)^3]\lineup =
E-\frac{1}{2}\lim_{\eps\to 0}\Tr[\Phi(\eps)\Gamma(\eps)]\nonumber\\
\lineup = \frac{1}{2\pi^2}\Big[-g(0)+(1+w)g(\infty)\Big].
\end{eqnarray}
which is incorrect.

\section{Ellwood Solutions}
\label{sec:Ellwood}

\subsection{Regularization}
\label{subsec:Ian_reg}

So far we have focused our analysis on the BMT solution. In a sense,
the BMT solution is a degenerate example of Ellwood's more general
construction, and it is important to understand how much our analysis
depends on accidental simplifications or singularities of this
particular example.

Therefore in this section we study Ellwood solutions. The first question
which arises is how the Ellwood solutions should be regularized. One
obvious approach is to define the regulated Ellwood solution as a gauge
transformation of the regulated BMT solution \eq{solution2}, by some
off-shell extension of the Zeze map \eq{Zeze}. However, this approach
leads to regularizations which
seem artificial, and it does not give any independent confirmation of the
physics behind the BMT solution. Furthermore, one of the most important
properties of Ellwood solutions is that they are (in general) much less
singular than the BMT solution from the perspective of the identity
string field. However, any gauge transformation of the BMT solution produces
a state which is essentially as identity-like as the BMT
solution.

Therefore we search for a way to regulate Ellwood solutions directly.
Compared to BMT, where the physics of the solution appears to be
reasonably independent of the choice of regularization, the regularization
of Ellwood solutions is delicate. Many natural proposals turn out to be
unphysical. We study one particular regularization which
works:\footnote{This regularization can be obtained as a particular
extension of the Zeze map:
$\Phi_{f,g}(\eps)
=f\Phi(\eps)\frac{1}{1+A\Psi_\mathrm{tv}(\eps)}g$.
Other naively equivalent possibilities, such as
$f\Phi(\eps)\frac{1}{1+A\Phi(\eps)}g$ and
$f\frac{1}{1+\Psi_\mathrm{tv}(\eps)A}\Phi(\eps)g$, appear not to work.}
the $\eps\to 0$ limit of the string field
\begin{equation}\Phi_{f,g}(\eps)
=f\left[c\phi-\frac{1}{K+\phi+\eps}\phi'\d c\right]
\frac{B}{1+\frac{1-fg}{K}(\phi+\eps)}c g.\label{eq:Ian_reg}\end{equation}
Under reasonable assumptions about $f$ and $g$, this is a well-defined
string field for all $\eps>0$. Like \eq{solution2}, this
expression can be separated into a tachyon vacuum solution
$\Psi^\mathrm{tv}_{f,g}(\eps)$ plus a term $\Delta_{f,g}(\eps)$ which
``builds'' the lump on top of the tachyon vacuum:
\begin{equation}\Phi_{f,g}(\eps)= \Psi^\mathrm{tv}_{f,g}(\eps)
+\Delta_{f,g}(\eps),\label{eq:Ell_PhiPsi}\end{equation}
where
\begin{eqnarray}\Psi^\mathrm{tv}_{f,g}(\eps)\lineup=
f\left[c(\phi+\eps)-\frac{1}{K+\phi+\eps}(\phi'+\eps)\d c\right]
 R(\eps)Bc g,\nonumber\\
\Delta_{f,g}(\eps)\lineup = f\left[-\eps c
+\frac{\eps}{K+\phi+\eps}\d c\right] R(\eps)Bc g,
\end{eqnarray}
and for short we have defined the factor
\begin{equation}R(\eps) \equiv \frac{1}{1+\frac{1-fg}{K}(\phi+\eps)}.
\label{eq:R}\end{equation}
Note the extra $\eps$ in the denominator of the rightmost factor of
\eq{Ian_reg}. This $\eps$ plays no role in regulating the Schwinger integral,
and if it was not there \eq{Ian_reg} would still be a well-defined string
field for $\eps>0$. Nevertheless this $\eps$ turns out to be necessary
to recover the correct physics in the infrared; it is not sufficient
to only regulate the divergent Schwinger integral. Also note that
\eq{Ian_reg} is not a gauge transformation of BMT. Therefore, to see
the correct physics emerge in the infrared requires genuinely new
calculations.

The Ellwood solution has the same problem with inverting $K+\phi$ as the
BMT solution, so we would not expect it to satisfy the
equations of motion. Computing the equations of motion from \eq{Ian_reg}
we find
\begin{equation} Q\Phi_{f,g}(\eps)+\Phi_{f,g}(\eps)^2 =
f\Big[\Gamma(\eps)-\Phi(\eps)\Delta(\eps)\Big]R(\eps)Bcg
+ \Big[f\Phi(\eps)R(\eps)Bcg\Big]\Big[f\Delta(\eps)R(\eps)Bc g\Big],
\end{equation}
where $\Phi(\eps),\Delta(\eps)$, and $\Gamma(\eps)$ are defined in
section \ref{sec:BMT}. The anomaly is a complicated expression, and we will
not attempt to calculate its overlap with test states. We do not expect it
to vanish in the $\eps\to 0$ limit.

\subsection{Overlap and Energy}
\label{subsec:Ian_ov}

We now show that the regularization \eq{Ian_reg} correctly
captures the physics of the lump, just like the BMT solution.

We start by computing the closed string overlap:
\begin{eqnarray}\Tr[\V\Phi_{f,g}(\eps)]\lineup
=\Tr[\V\Phi(\eps)R(\eps)Bc\,\omega]\nonumber\\
\lineup =\Tr[\V c\phi R(\eps)Bc\,\omega]
-\Tr\left[\V
\frac{1}{K+\phi+\eps}\phi'\d c R(\eps)Bc\,\omega\right], \label{eq:Ellov1}
\end{eqnarray}
where $\omega\equiv fg$ for short. Now in the second term insert
a trivial factor of $Bc$ next to the $\d c$. This allows us to eliminate the
$Bc$ between $R(\eps)$ and $\omega$ to find
\begin{equation}\Tr[\V\Phi_{f,g}(\eps)]=\Tr[\V c\phi R(\eps)Bc\,\omega]
-\Tr\left[\V
B\phi'c\d c\left( R(\eps)\,\omega\frac{1}{K+\phi+\eps}\right)\right].
\end{equation}
The factor in parentheses on the right can be simplified in a
useful way. To see how, we write $\omega$ in a particular form:
\begin{eqnarray}\omega \lineup = -\frac{1-\omega}{K}K+1,\nonumber\\
\lineup = -\frac{1-\omega}{K}(K+\phi+\eps)+1+\frac{1-\omega}{K}(\phi+\eps),
\nonumber\\
\lineup = R(\eps)^{-1}-\frac{1-\omega}{K}(K+\phi+\eps).
\end{eqnarray}
Therefore
\begin{equation}R(\eps)\,\omega\frac{1}{K+\phi+\eps}=\frac{1}{K+\phi+\eps}
-R(\eps)\frac{1-\omega}{K}.
\label{eq:Ellov3}\end{equation}
Plugging this into the parentheses of \eq{Ellov1} the $\frac{1}{K+\phi+\eps}$
term above combines with the insertions to give the overlap of the
regularized BMT solution. Keeping track of the other terms gives
\begin{equation}\Tr[\V\Phi_{f,g}(\eps)]=\Tr[\V\Phi(\eps)]
+\Tr\Big[\V c\phi R(\eps)Bc\,\omega\Big]
+\Tr\left[\V B\phi'c\d c R(\eps)\frac{1-\omega}{K}\right].
\end{equation}
The overlap of the BMT solution was already computed in section
\ref{subsec:BMT_reg} and \cite{BMT}, and we know it gives the correct
shift in the closed string tadpole. All we have to do is show that the
last two terms cancel in the $\eps\to 0$ limit. This can be
done with a few specially chosen manipulations:
\begin{eqnarray}\Tr[\V\Phi_{f,g}(\eps)]-\Tr[\V\Phi(\eps)]\lineup =
\Tr\left[\V R(\eps)Bc\left(\omega c\phi
+B\frac{1-\omega}{K}\phi'c\d c\right)\right],\nonumber\\
\lineup=-\Tr\left[\V R(\eps)Bc\,Q\left(B\frac{1-\omega}{K}c\phi\right)\right].
\end{eqnarray}
Replace the $R(\eps)Bc$ factor in the trace by an equivalent
expression dressed up with redundant factors of $Bc$:
\begin{eqnarray}\Tr[\V\Phi_{f,g}(\eps)]-\Tr[\V\Phi(\eps)]\lineup =
-\Tr\left[\V \frac{1}{1+B\frac{1-\omega}{K}c(\phi+\eps)}
Q\left(B\frac{1-\omega}{K}c\phi\right)\right].
\end{eqnarray}
The right hand side is almost the overlap of a pure gauge
solution. We just have to fix up the string field inside the BRST variation:
\begin{eqnarray}\Tr[\V\Phi_{f,g}(\eps)]-\Tr[\V\Phi(\eps)]\lineup =
-\Tr\left[\V \frac{1}{1+B\frac{1-\omega}{K}c(\phi+\eps)}
Q\left(1+B\frac{1-\omega}{K}c(\phi+\eps)\right)\right],\nonumber\\
\lineup\ \ \ \ \ \
+\eps\Tr\left[\V R(\eps)Bc\,Q\left(B\frac{1-\omega}{K}c\right)\right].
\end{eqnarray}
The first term is the closed string overlap of a pure gauge solution.
Assuming $R(\eps)$ admits a convergent geometric series expansion in powers
of $\frac{1-\omega}{K}(\phi+\eps)$, this term can be shown to vanish
order by order. Therefore only the second term contributes for finite $\eps$,
giving
\begin{equation}\Tr[\V\Phi_{f,g}(\eps)]=\Tr[\V\Phi(\eps)]
+\eps\Tr\left[\V R(\eps)Bc\,Q\left(B\frac{1-\omega}{K}c\right)\right].
\label{eq:Ellov2}\end{equation}
The second term vanishes since the string field in the
trace is regular in the $\eps\to 0$ limit. Therefore the Ellwood
solutions \eq{Ian_reg} have precisely the correct coupling to closed string
states.

The overlaps of the Ellwood and BMT solutions are not precisely equal for
$\eps>0$ because the regularized Ellwood solution \eq{Ian_reg} is not a
gauge transformation of the regularized BMT solution \eq{solution2}.
However, one might think that the overlaps must be equal in the
$\eps\to 0$ limit because the Ellwood and BMT solutions are (formally)
gauge equivalent. Actually, this is not the case. Many other
regularizations of the Ellwood solution do not reproduce the correct
coupling to closed strings. Consider what would happen if we had only
regulated the Schwinger integral, and not included the extra $\eps$
in the rightmost factor of \eq{Ian_reg}. In this case the regularized
Ellwood solution would be
\begin{equation}\Phi_{f,g}(\eps) = f\Phi(\eps)R(0)Bc g,\end{equation}
where $R(0)$ is \eq{R} at $\eps=0$. The calculation of the overlap
proceeds analogously, but, to extract the overlap of the regularized BMT
solution, instead of \eq{Ellov3} we need the relation
\begin{equation}R(0)\,\omega\frac{1}{K+\phi+\eps}=\frac{1}{K+\phi+\eps}
-R(0)\frac{1-\omega}{K}-R(0)\frac{1-\omega}{K}\frac{\eps}{K+\phi+\eps}.
\end{equation}
The main difference between this and \eq{Ellov3} is the third term. While
the third term is proportional to $\eps$, it is nonvanishing in the
$\eps\to 0$ limit, and actually gives the sole contribution to the difference
between the Ellwood and BMT overlaps:
\begin{equation}\Tr[\V\Phi_{f,g}(\eps)]=\Tr[\V\Phi(\eps)]
+\Tr\left[\V B\phi'c\d c R(0)\frac{1-\omega}{K}\frac{\eps}{K+\phi+\eps}
\right].\end{equation}
In the $\eps\to 0$ limit the second term becomes
\begin{equation}
\Tr\left[\V B\phi'c\d c R(0)\frac{1-\omega}{K}\tilde{\Omega}^\infty\right].
\end{equation}
This does not appear to vanish for generic choice
of $f$ and $g$. Therefore the fact that the closed string overlap works
in \eq{Ellov2} is not a consequence of a formal gauge equivalence, but
is an independent confirmation of the physics behind the construction.

We can calculate the energy of the Ellwood solutions \eq{Ian_reg}
by analogy with the BMT solution. The idea is to extract the negative
energy from the tachyon
vacuum, and to reduce the remaining terms to their BMT counterparts
by repeated use of the identity \eq{Ellov3}. The calculation requires keeping
track of many terms, and is too lengthy and mostly routine to be
worth presenting here. Some aspects however deserve mention.
The first is that for the Ellwood solutions
(with this regularization) there is no simple relation between
the anomaly, solution, and phantom term analogous to \eq{dpg}. This relation
was crucial for simplifying the action from \eq{step1} to \eq{step2}.
For Ellwood solutions this simplification does not happen automatically,
and the terms which would otherwise simplify have to be expanded and
shown to cancel in a nontrivial fashion. The second point is that the
calculation produces many spurious terms which do not cancel identically for
$\eps>0$. Most of these terms are impractical to explicitly compute for
$\eps>0$, and they must be argued to vanish for general reasons in the
$\eps\to 0$ limit. These terms take one of three forms:
\begin{equation}\eps\Tr[X(\eps)],\ \ \ \ \ \
\eps\Tr\left[X(\eps)\frac{\eps}{K+\phi+\eps}
\right],\ \
\ \ \ \ \Tr\left[X(\eps)B\frac{\eps}{K+\phi+\eps}\right],
\end{equation}
where $X(\eps)$ is a finite and not sliver-like string field, generally
some combination of $R(\eps),\ \omega,\ \frac{1-\omega}{K},\ \phi,\ \phi'$,
and ghosts. The first two classes of terms vanish because
an overall factor $\eps$ multiplies a trace we believe is finite in
the $\eps\to 0$ limit. The third class of terms vanish because in the
$\eps\to0$ limit the ghost component of the correlator has insertions with
effectively positive scaling dimension on a very large cylinder. This is
essentially the reasons why $B$ annihilates the sliver in the Fock space.
With this understanding, the calculation of the energy is straightforward
and reproduces the expected answer \eq{energy}.

\section{Cohomology?}
\label{sec:cohomology}

It is interesting to ask whether the BMT solution supports the
expected cohomology of open string states. Of course, taken literally this
question has no meaningful answer, since the shifted kinetic operator is not
nilpotent:
\begin{equation}Q_{\Phi(\eps)}^2=[\Gamma(\eps),\cdot]\label{eq:Q2}
\end{equation}
So the existence of cohomology is closely related to the equations
of motion.  In this section we argue that the BMT solution satisfies the
equations of motion when contracted with states of the IR boundary conformal
field theory, in a sense described below. Then, the BMT kinetic operator is
nilpotent in $\mathrm{BCFT}^*$, and defines
a cohomology.

As a first step, let us explain what it means to contract the equations of
motion, which is a state in $\mathrm{BCFT}_0$, with states in
$\mathrm{BCFT}^*$. Suppose
\begin{equation}\Pi_i^* = \Omega^{1/2}\pi_i^*\Omega^{1/2}\end{equation}
are a basis of Fock states of $\mathrm{BCFT}^*$, where
$\pi_i^*$ are vertex
operators and (in this equation) $\Omega$ is the $SL(2,\mathbb{R})$ vacuum of
$\mathrm{BCFT}^*$. This basis can be equivalently characterized in
$\mathrm{BCFT}_0$ as a singular, projector-like limit of states of the
form
\begin{equation}\Pi_i(\eps) = \tilde{\Omega}^{1/2\eps}\pi_i(\eps)
\tilde{\Omega}^{1/2\eps}, \label{eq:PsiiL}\end{equation}
when $\eps\to 0$. The equivalence between $\Pi_i^*$ and $\Pi_i(\eps)$
can be explained as follows: In correlation functions on the cylinder,
$\Pi_i(\eps)$ represents a strip of worldsheet of width $1/\eps$ with deformed
boundary conditions and an operator $\pi_i(\eps)$ inserted in the middle.
With a reparameterization, we can squeeze the strip to width $1$,
whereupon the boundary conditions flow to $\mathrm{BCFT}^*$, and the operator
$\pi_i(\eps)$, if appropriately
chosen\footnote{We will not attempt here to construct the full
basis of states $\Pi_i(\eps)$ explicitly for a particular relevant
deformation, though we have studied a few examples. However, a few points
are worth mentioning. First, the operators
$\pi_i(\eps)$ are not fixed uniquely.
For example, for the Witten deformation, both $\eps c$ and
$-\frac{\eps}{u}c\phi$ flow to the zero momentum tachyon $c$ in the
infrared. Second, many operators, such as the energy momentum tensor,
experience divergent contractions with the boundary interaction and must be
appropriately renormalized. Lastly, the states $\Pi_i(\eps)$ in general diverge
in the Fock space of $\mathrm{BCFT}_0$ in the $\eps\to 0$ limit.}, flows to
$\pi_i^*$. In particular, this means that $N$-string vertices of
$\Pi_i(\eps)$, when $\eps\to0$, are equal to the corresponding $N$-string
vertices of $\Pi_i^*$, and so for string field theory purposes the states are
indistinguishable.

With this understanding, the BMT solution satisfies the equations
of motion in $\mathrm{BCFT}^*$ in the following sense:
\begin{equation}
\lim_{\eps\to 0}\Big\langle \Pi_i(\eps),\,
Q\Phi(\eps)+\Phi(\eps)^2\Big\rangle
=\lim_{\eps\to 0}\Big\langle \Pi_i(\eps),\Gamma(\eps)\Big\rangle=0.
\label{eq:anzero}\end{equation}
To see why \eq{anzero} holds, note that, because of the
very large width of the test state $\Pi_i(\eps)$, contractions between
$\phi'$ and the vertex operator $\pi_i(\eps)$ are suppressed by cluster
decomposition. Therefore
\eq{anzero} should be proportional to the one point function of $\phi'$ on a
very large (deformed) cylinder. This
vanishes {\it faster} than $\eps$ because the disk partition
function is constant in the infrared, by \eq{phipdL}. The ghost correlator
diverges as $\frac{1}{\eps}$, but this is not enough to cancel the vanishing
matter correlator.

Let us clarify a possibly confusing point: \eq{anzero} does not imply that
the BMT solution is a well defined state in $\mathrm{BCFT}^*$
satisfying the equations of motion. In fact it is not: the overlap of
$\Phi(\eps)$ with $\Pi_i(\eps)$ diverges in the $\eps\to 0$ limit. However,
the anomaly in the equations of motion {\it is} a well-defined state in
$\mathrm{BCFT}^*$, and it is precisely zero. This is all we need for the
cohomology.

Acting within $\mathrm{BCFT}^*$, the BMT kinetic operator takes the form
\begin{equation}\Big\langle \Pi_i(\eps),\,Q_{\Phi(\eps)}\Pi_j(\eps)\Big\rangle
 =
\Big\langle \Pi_i(\eps),\,Q\Pi_j(\eps)+[\Phi(\eps),\Pi_j(\eps)]\Big\rangle.
\label{eq:BMTQIR}\end{equation}
Since the BMT solution is not well-defined in $\mathrm{BCFT}^*$, it is not
obvious that the BMT kinetic operator should be meaningful either.
To see what happens, concentrate first on the action
of the BRST operator, $Q$. When $Q$ acts on a deformed wedge state, it can
be naturally separated into two pieces:
\begin{equation}Q = -[c\phi,\,\cdot\,]+\tilde{Q}.\label{eq:Qdecomp}
\end{equation}
If $\phi$ were a marginal operator, the first part would be the
BRST variation of the boundary condition changing operator, and the second
part, $\tilde{Q}$, would be the BRST operator of the marginally deformed
boundary conformal field theory. Since $\phi$ is not marginal,
$\tilde{Q}$ is not a BRST charge, and it is not nilpotent:
\begin{equation}\tilde{Q}^2=[\phi'c\d c,\,\cdot\,].\end{equation}
However, the operator $\phi'c\d c$ vanishes in $\mathrm{BCFT}^*$ for the
same reason that the anomaly vanishes. Therefore, in the $\eps\to 0$ limit
$\tilde{Q}$ is nilpotent and can be
naturally identified with the BRST operator of the infrared boundary
conformal field theory:
\begin{equation}\lim_{\eps\to 0}\langle \Pi_i(\eps),\tilde{Q}\Pi_j(\eps)
\rangle
= \langle \Pi_i^*,Q\Pi_j^*\rangle.\end{equation}
The $-[c\phi,\cdot]$ term in the BRST operator diverges as
$\eps\to 0$, but thankfully it cancels against the corresponding divergence
from the BMT solution in \eq{BMTQIR}. The remaining piece of the BMT solution,
$-\frac{1}{K+\phi+\eps}B\phi'c\d c$, does not contribute in the $\eps\to0$
limit because the $1$-point function of $\phi'$ kills the matter correlator.
Adding everything up gives the simple result:
\begin{equation}\lim_{\eps\to 0}\Big\langle \Pi_i(\eps),Q_{\Phi(\eps)}
\Pi_j(\eps)\Big\rangle
= \langle \Pi_i^*,Q\Pi_j^*\rangle.\end{equation}
Therefore the BMT kinetic operator in the infrared is the same as the BRST
operator of $\mathrm{BCFT}^*$, and they share the same cohomology.

It is interesting to see how the cohomology disappears for the tachyon vacuum
regularization of the BMT solution, $\Psi_\mathrm{tv}(\eps)$ in \eq{tv}.
Repeating the above steps, in the tachyon vacuum
kinetic operator in the IR takes the form
\begin{equation}\lim_{\eps\to0}\Big\langle \Pi_i(\eps),
Q_{\Psi_\mathrm{tv}(\eps)}
\Pi_j(\eps)\Big\rangle
= \langle \Pi_i^*,Q\Pi_j^*\rangle
-\lim_{\eps\to 0}\Big\langle\Pi_i(\eps),[\Delta(\eps),\Pi_j(\eps)]\Big\rangle,
\end{equation}
Performing a scale transformation of the second term, we can replace
$\Pi_i(\eps)$ with $\Pi_i^*$ and $\Delta(\eps)$ with
$-c+\frac{1}{K+1}Bc\d c$, where ``$K$'' is now the $K$
of the IR boundary conformal field theory. Therefore
\begin{equation}\lim_{\eps\to 0}\Big\langle \Pi_i(\eps),
Q_{\Psi_\mathrm{tv}(\eps)}
\Pi_j(\eps)\Big\rangle
= \left\langle \Pi_i^*,Q\Pi_j^*+\left[\frac{1}{K+1}(c+Q(Bc)),\Pi_j^*\right]
\right\rangle.
\end{equation}
The right hand side is precisely the kinetic operator of the ``simple''
tachyon vacuum solution described in \cite{simple}. Note that because
\begin{equation}\lim_{\eps\to 0}\Big\langle \Pi_i(\eps),\Delta(\eps)
\Big\rangle
=- \left\langle \Pi_i^*,\frac{1}{K+1}(c+Q(Bc))
\right\rangle,
\label{eq:Dsimple}\end{equation}
$\Delta(\eps)$ is a well defined state in $\mathrm{BCFT}^*$, and
is precisely the perturbative vacuum of $\mathrm{BCFT}^*$ as seen from the
tachyon vacuum. This is consistent with the interpretation of $\Delta(\eps)$
as ``building'' the lump on top of the tachyon vacuum.

\section{Concluding Remarks}

In this paper we studied a class of formal solutions which were conjectured
to describe lower dimensional branes as tachyon lumps in open string field
theory. We found that the solutions do
not satisfy the equations of motion. Nevertheless,
they have the correct coupling to closed string states, and
evaluating the action gives the expected energy.

The current situation is puzzling since the correct solution
remains to be found, yet clearly this construction is capturing the
physics of the desired solution in a nontrivial fashion. The question
now is how to proceed. We offer a few possibilities:
\begin{itemize}
\item It is possible that while the specific solutions \eq{Ian_sol} are
problematic, other solutions within the subset of states generated by
multiplying $K,B,c,\phi$, and $\phi'$ could describe a tachyon lump. It is
difficult to analyze the full set of candidate solutions in
generality, but we have found that the difficulties with Ellwood's
proposal are fairly generic. Perhaps a novel
mechanism selects a particular subclass of solutions for which the equations
of motion can be made non-anomalous.
\item It is possible that the Ellwood and BMT
solutions satisfy the equations of motion when correctly defined,
but we have not identified the
necessary definition of expressions such as $\frac{1}{K+\phi}$ when they
appear inside the solution. It is worth mentioning that analogous problems
with
defining $\frac{1}{K}$ appear when studying of multibrane
solutions\cite{multiple}, and new developments on this front may also have
implications for lump solutions.
\item Finally, it is possible that the current setup is for some reason
inadequate to capture nonsingular lump solutions. Perhaps a
different approach, for example based on boundary condition changing
operators, as suggested in \cite{simple_marg}, is needed.
\end{itemize}
We hope that the current work will stimulate further thought on this important
problem.

\bigskip
\noindent {\bf Acknowledgments}
\bigskip

\noindent We would like to thank the organizers of the conference SFT 2010 in Kyoto where this collaboration began, and Micheal Kiermaier , Yuji Okawa, and Martin Schnabl for useful conversations.
We would like to thank L. Bonora for making  his  numerical computations available to us. 
We thank Ian Ellwood for comments on the second version of the paper.
This research was supported by the EURYI grant
GACR EYI/07/E010 from EUROHORC and ESF.

\begin{appendix}

\section{Witten Deformation}
\label{app:Witten}

In this appendix we give some formulas which allow for explicit computation
of correlation functions on the cylinder in the presence of the Witten
boundary interaction. Most of the formulas follow immediately from
\cite{Witten} with the appropriate transcription.

The Witten deformation is generated by inserting the operator
\begin{equation}\exp\left[-\int_0^L ds\,\phi_u(s)\right]\end{equation}
into correlation functions on the cylinder in a reference $\mathrm{BCFT}_0$
which includes a noncompact free boson $X(z,\bar{z})$ subject to Neumann
boundary conditions, where
\begin{equation}
\phi_u(s)=u\left[\frac{1}{2}:X(s)^2:+\gamma-1+\ln(2\pi u)\right]\\
\end{equation}
and
\begin{equation}
\phi'_u(s)=u\frac{d}{du}\phi_u(s)=\phi_u(s)+u.
\end{equation}
Here $u$ is a parameter which we are free to choose. Different $u$s are
related by the scale transformation \eq{phi_u}. For short, let's write
\begin{equation}\langle ... \rangle^{u}_{C_L}
=\left\langle\exp\left[-\int_0^L
ds\,\phi_u(s)\right]...\right\rangle_{C_L}^{\mathrm{BCFT}_0}.\end{equation}
The $X$ part of the worldsheet theory is Gaussian and therefore completely
defined by the zero point and bulk 2-point functions
\begin{eqnarray}\langle 1\rangle^{u}_{C_L}\lineup =g(L)\\
\langle X(z_1,\bar{z}_1)X(z_2,\bar{z}_2)\rangle^{u}_{C_L}\lineup
=g(L){\bf G}(z_1,\bar{z_1};z_2,\bar{z}_2;L),
\end{eqnarray}
where, leaving the $u$ dependence implicit, $g(L)$ is the disk partition
function and ${\bf G}$ is the bulk Green's function:
\begin{eqnarray}g(L)\lineup
=\sqrt{\frac{uL}{2\pi}}\left(\frac{e}{uL}\right)^{uL}
\Gamma(uL)\\
{\bf G}(z_1,\bar{z_1};z_2,\bar{z}_2;L)\lineup =
-\frac{1}{2}\left(\ln|Z_1-Z_2|^2-\ln|1-Z_1\bar{Z}_2|^2\right)\nonumber\\
\lineup\ \ \ \ \ \ \ \ \ \ \ \ \ \ \ \ \
-\frac{1}{uL}+2\mathrm{Re}\big[\Phi(Z_1\bar{Z_2},1,uL)\big]\ \ \ \ \ \ \ \
(Z=e^{\frac{2\pi i}{L}z}),\ \ \ \ \ \ \ \
\end{eqnarray}
where $\Phi(z,s,a)$ is the Lerch zeta function
\begin{equation}\Phi(z,s,a)=\sum_{k=0}^\infty \frac{z^k}{(k+a)^s}.\end{equation}

For most applications we are interested in computing $n$-point functions of
$\phi_u(s)$. For this purpose it is helpful to define the boundary Green's
function and the normalized 1-point function:
\begin{eqnarray}G(s,L)\lineup = -\frac{1}{uL}
+2\mathrm{Re}\left[\Phi\left(e^{\frac{2\pi i}{L}s},1,uL\right)\right],
\label{eq:G_boundary}\\
W(L)\lineup = -u\left[1+\frac{1}{2uL}+\psi(uL)-\ln(uL)\right],\label{eq:W}
\end{eqnarray}
where $\psi$ is the digamma function. With these objects can write
explicit expressions for the $0,1,2$ and $3$ point functions using Wick's
theorem:
\begin{eqnarray}\langle 1\rangle^u_{C_L}\lineup =g(L),\\
\langle\phi_u(s)\rangle^u_{C_L}\lineup = g(L)W(L),\\
\langle\phi_u(s_1)\phi_u(s_2)\rangle^u_{C_L}
\lineup =g(L)\left[W(L)^2+\frac{u^2}{2}G(s_{12},L)^2\right],\label{eq:2-point}\\
\langle\phi_u(s_1)\phi_u(s_2)\phi_u(s_3)\rangle^u_{C_L}\lineup =
g(L)\left[W(L)^3+\frac{u^2}{2}W(L)\big(G(s_{12},L)^2+G(s_{23},L)^2
+G(s_{31},L)^2\big)\right.\nonumber\\
\lineup \ \ \ +u^3G(s_{12},L)G(s_{23},L)G(s_{31},L)\Big].\label{eq:3-point}
\end{eqnarray}
We can write similar expressions for higher point functions as
well, but we will not need them.

For computations related to the anomaly it is useful to have asymptotic
formulas for the large $L$ behavior of these $n$-point functions.
For this purpose we list some large $L$ expansions:
\begin{eqnarray}g(L)\lineup =1\,+\,\frac{1}{12}\frac{1}{uL}
\,+\,\frac{1}{288}\frac{1}{(uL)^2}\,
-\,\frac{139}{51840}\frac{1}{(uL)^3}+...\nonumber\\
\lineup =
\exp\left[\sum_{n=1}^\infty\frac{B_{2n}}{2n(2n-1)}\frac{1}{(uL)^{2n-1}}\right],
\\
W(L)\lineup
=-u\left[\,1\,-\,\frac{1}{12}\frac{1}{(uL)^2}\,
+\,\frac{1}{120}\frac{1}{(uL)^4}\,-\,...
\right]\nonumber\\
\lineup = -u\left[1-\sum_{n=1}^\infty\frac{B_{2n}}{2n}\frac{1}{(uL)^{2n}}
\right],
\end{eqnarray}
where $B_n$ are the Bernoulli numbers. For the boundary Green's function we
can derive an asymptotic expansion for large $L$ and fixed separation
 $s$ between the $X$ insertions:
\begin{equation}G(s,L)=
-2\Big[\cos(2\pi u s)\ci(2\pi u s)+\sin(2\pi us)\si(2\pi us)\Big]
+2\sum_{n=1}^\infty \frac{B_{2n}}{2n (uL)^{2n}}\cos_{n}(2\pi us),
\label{eq:fixed_s}
\end{equation}
where $\si$ and $\ci$ are the sine and cosine integrals \eq{sici} and
$\cos_{n}$ is a partial sum of the cosine series,
\begin{equation}\cos_{n}(x)=\sum_{k=0}^{n-1}\frac{(-1)^k x^{2k}}{(2n)!}.
\end{equation}
Another useful formula is the large $L$, fixed $q=\frac{s}{L}$ behavior of
the boundary Green's function\footnote{We start from the asymptotic
formula appearing in equation (7) of \cite{Lerch}. This formula corrects
equation (D.24) in \cite{BMT}.}
\begin{eqnarray}G(qL,L) \lineup = \frac{1}{(uL)^2}\frac{1}{2}\csc^2(\pi q)
\,-\,\frac{1}{(uL)^4}\frac{1}{4}\csc^4(\pi q)\big[2+\cos (2\pi q)\big]\,+\,...
\nonumber\\ \lineup = 2\sum_{n=0}^\infty (-1)^n\left(\frac{\csc(\pi q)}{2uL}
\right)^{2n+2}\sum_{k=0}^{2n} \left\langle{2n+1 \atop k}\right\rangle
\cos\big[2\pi(k-n)q\big],\label{eq:fixed_Q}
\end{eqnarray}
where $\left\langle {m\atop n}\right\rangle$ are the Eulerian numbers. Note
that, in this expansion, the boundary Green's function vanishes as $1/L^2$
for large $L$, as we would expect since the boundary conditions on the
cylinder become Dirichlet ($X=0$) in the $L\to\infty$ limit. However,
the leading $1/L^2$ behavior comes with a coefficient which depends on the
normalized separation $q$ between the $X$ insertions which has a double
pole when the insertions become coincident. This is not the usual logarithmic
behavior we would expect from the $X$-$X$ OPE, and the double pole
accounts for the fact that the large $L$ but fixed $s$ limit of $G(s,L)$
\eq{fixed_s} is actually nonzero despite the fact that the boundary
conditions are becoming Dirichlet as $L\to\infty$. This is essentially the
reason why the anomaly in the equations of motion \eq{anomaly} does not
vanish.

\section{Tachyon Vacuum Energy for Witten Deformation}
\label{app:vac_energy}

In this appendix we compute the action of the tachyon vacuum solution
\eq{tv} for the Witten deformation. Our computation serves as an
independent check of the effectively equivalent numerical computation first
appearing in \cite{B}. Thanks to improved analytic control of the
Green's functions, we are able to obtain more precision.

Using the results of equations \eq{D3} and \eq{PhiG}, we can compute the
action of the tachyon vacuum in terms of the cubic vertex evaluated on the
BMT solution:
\begin{equation}\Tr\left[\frac{1}{6}\Psi_\mathrm{tv}(\eps)^3\right]=
\lim_{\eps\to 0}\Tr\left[\frac{1}{6}\Phi(\eps)^3\right]
+\frac{g(\infty)}{2\pi^2}(1+w).\label{eq:n_tv}\end{equation}
Focus on the computation of $\Tr[\Phi(\eps)^3]$.
Substituting the solution, expanding out the Schwinger integrals, and
evaluating the matter and ghost correlators
gives an expression of the form
\begin{equation}\lim_{\eps\to 0}\Tr\left[\frac{1}{6}\Phi(\eps)^3\right]
=-\frac{1}{6}\lim_{\ell\to 0}\int_\ell^\infty dL \int_0^1 dq
\int_0^{1-q} dr\, F(L,q,r)K(q,r).\label{eq:int1}\end{equation}
We have already taken the $\eps\to 0$ limit on the right hand side, since the
integration is convergent for large $L$. However, we introduce a
unrelated regularization $\ell\to 0$ for the lower limit of the integral. This
has nothing to do with singularities of the BMT solution, but represents
a regularization of the divergent energy from the infinite
volume of the reference D-brane. We will say more about this in a moment.
The functions $F$ and $K$ above come from evaluating the appropriate
matter/ghost correlators. The ghost factor is given by
\begin{equation}
K(q,r)=-\frac4\pi \,\sin\pi q\,\sin\pi r\,
\sin\pi(q+r),
\end{equation}
and the matter factor comes from the 3-point function of $\phi'$, which
can be written as the sum of three terms:
\begin{equation}
 F(L,q,r)=F_1(L)+F_2(L,q,r)+F_3(L,q,r),\label{eq:three}
\end{equation}
where
\begin{eqnarray}
F_1(L)\lineup = L^2g(L)\Big(W(L)+1\Big)^3,\\
F_2(L,q,r)\lineup=
\frac1{2}L^2g(L)\Big(W(L)+1\Big)
\Big(G(Lq,L)^2+G(L(r+q),L)^2+G(L r,L)^2\Big) ,\\
F_3(L,q,r)\lineup =L^2g(L)\Big(G(L q,L)G(L(r+q),L)G(L r,L)\Big),
\end{eqnarray}
The integral \eq{int1} is independent of $u$ and we are
free to choose a canonical value. We set $u=1$.

The matter correlator $F(L,q,r)$ diverges as $\frac{1}{\sqrt{L^3}}$ for
small $L$, and therefore the integral \eq{int1}
diverges as $\frac{1}{\sqrt{\ell}}$ in the $\ell\to 0$ limit. This divergence
is related to the divergence of the norm
of the $SL(2,\mathbb{R})$ vacuum in $\mathrm{BCFT}_0$, which corresponds
to the $\ell\to 0$ limit of the expression
\begin{equation}g(\ell)=\frac{1}{\sqrt{2\pi}}
\frac{1}{\sqrt{\ell}}
+\mathcal{O}(\sqrt{\ell}).\label{eq:gexp}\end{equation}
This suggests that the integral \eq{int1} can be defined by subtracting the
$\frac{1}{\sqrt{\ell}}$ divergence and replacing it with
$\frac{g(0)}{2\pi^2}$.\footnote{The numerical factor in front of
$\frac{1}{\sqrt{\ell}}$ is irrelevant, up to an overall sign, since it can
be absorbed into a redefinition $\ell\to constant\times\ell$, and
at any rate $\ell$ is going to zero.}
Note that because the subleading terms in \eq{gexp} vanish as $\ell\to 0$,
we should subtract {\it only} the $\frac{1}{\sqrt{\ell}}$ divergence and leave
the finite remainder untouched. Therefore we write
\begin{equation}\lim_{\eps\to 0}\Tr\left[\frac{1}{6}\Phi(\eps)^3\right]
=\frac{g(0)}{2\pi^2}-\frac{1}{6}\int_0^\infty dL \int_0^1 dq
\int_0^{1-q} dr\, F^*(L,q,r)K(q,r).\label{eq:int2}\end{equation}
The function $F^*(L,q,r)$ is related to $F(L,q,r)$ by adding a total
derivative term:
\begin{equation}
F^*(L,q,r) = F(L,q,r)
+\frac{15}{8\sqrt{2\pi}}\frac{d}{dL}\left(\frac{1}{L^{1/2}}e^{-L}\right).
\label{eq:total_der}\end{equation}
The coefficient in front of the total derivative has been fixed so that
integration produces a boundary term at $L=\ell$ which precisely
cancels the $\frac{1}{\sqrt{\ell}}$ divergence
from $F$. The finite contribution from $F$ is unchanged because
the subleading contributions from the boundary term at $L=\ell$ vanish as
$\ell\to 0$, and the $L=\infty$ boundary term vanishes due
to the $e^{-L}$ suppression. Actually, in the following we will compute the
contribution to the energy from the three terms $F_1,F_2$ and $F_3$
separately. Accordingly, we define subtracted functions $F_1^*,F_2^*$ and
$F_3^*$ following the above prescription.

With this preparation, we can put these integrals into a computer.
The $F_1$ integral is easily done since only the ghost sector
enters into the integration over $q$ and $r$, which can be performed
analytically. The remaining numerical integral over $L$ gives the
result
\begin{equation}
-\frac16\int_0^\infty dL\int_0^1dq\int_0^{1-q} dr\,K(q,r)\,
F_1^*(L)=-\frac1{2\pi^2}(0.406818).
\end{equation}
To evaluate the contribution from $F_2$ we observe that because of a
symmetry $q\to1-q$ of the Green's function, $G(Lq,L)$, we can replace
\begin{equation}
F_2(L,q,r)\rightarrow \frac{3}{2}L^2\Big(W(L)+1\Big)G(L q,L)^2.
\end{equation}
inside the integral. The integral over $r$ now only involves the ghost factor
$K(q,r)$, and the remaining integration over $q$ and $L$ can be done
numerically. (To help the computer in the $L\sim0$ region we found
it convenient to substitute the integrand with its first term in the
Taylor expansion in the interval  $L\in(0,10^{-9})$, so that the
$q$-integration can be performed exactly in this region. This introduces a
small error which, as we checked, is under control and of order $10^{-5}$).
In total
\begin{equation}
-\frac16\int_0^\infty dL\int_0^1dq\int_0^{1-q} dr\,K(q,r)\, F_2^*(L,q,r)
=-\frac1{2\pi^2}(0.9862)\,.
\end{equation}
To compute the final contribution from $F_3$ we use a
trick to get rid of one integral analytically. Note that $K(q,r)F_3(L,q,r)$
can written as the product of three copies of a single function evaluated
at $q$, $r$ and $q+r$. Inserting the appropriate step
functions we can extend
the range of integration over $q$ and $r$ from plus to minus infinity, and
inserting an auxiliary integral over $s$ (with a delta function) allows \
us to write
\begin{equation}
\int_0^1 dq\int_0^{1-q}K(q,r)F_3(L,q,r)= -\frac{4}{\pi} L^2 g(L)
\int_{\mathbb{R}^3}dqdrds\, \delta(q+r-s)h(L,q)h(L,r)h(L,s),
\label{eq:cub_vert}\end{equation}
where
\begin{equation}
h(L,q)=\sin(\pi q)G(L q,L)\theta_{[0,1]}(q),
\end{equation}
and $\theta_{[0,1]}(q)$ is a unit step function with support on the interval
$q\in[0,1]$. Note that \eq{cub_vert} looks like the cubic vertex of a field
$h$ in momentum space, with a delta function for momentum conservation.
With a Fourier transform, the three integrals over the momenta
turn into a single integral over the interaction point $x$:
\begin{equation}
\int_0^1 dq\int_0^{1-q}K(q,r)F_3(L,q,r)= -\frac{2}{\pi^2} L^2 g(L)
\int_{\mathbb{R}}dx\,\tilde{h}(L,x)\left|\tilde{h}(L,x)\right|^2,
\end{equation}
where
\begin{equation}
\tilde{h}(L,x)=\int_\mathbb{R} dq\,h(L,q)\,e^{i qx},
\end{equation}
The function $\tilde h(L,x)$ can be computed analytically, although its form
is not particularly interesting to write it down. Therefore, to compute
the $F_3$ contribution we only need to evaluate a numerical integral over
over $L$ and $x$. This gives the result
\begin{equation}
-\frac16\int_0^\infty dL\int_0^1
dq\int_0^{1-q} dr\,K(q,r)\, F_3^*(L,q,r)=-\frac1{2\pi^2}(-0.0263029).
\end{equation}
Adding the contributions from $F_1$, $F_2$, and $F_3$ together, we find
\begin{eqnarray}
\lim_{\eps\to0}\Tr\left[\frac{1}{6}\Phi(\eps)^3\right]\lineup
=\frac{g(0)}{2\pi^2}-
\frac{g(\infty)}{2\pi^2}(0.406818+0.9862-0.0263029),\nonumber\\
\lineup =\frac{g(0)}{2\pi^2}-\frac{g(\infty)}{2\pi^2}(1.3668).
\end{eqnarray}
Plugging into \eq{n_tv}, the terms proportional to $g(\infty)$ cancel within
the expected error, leaving the energy for the tachyon
vacuum solution:
\begin{equation}
E=-\frac{1}{6}\Tr[\Psi_\mathrm{tv}(\eps)^3]=-\frac{g(0)}{2\pi^2}.
\end{equation}

\end{appendix}

\end{document}